\documentclass[11pt]{article}

\usepackage{amssymb,amsfonts,amsmath,amsthm,amscd,dsfont,mathrsfs}
\usepackage{graphicx,float,psfrag,epsfig}
\usepackage{wrapfig}
\usepackage{relsize}
\usepackage{color}
\usepackage{pict2e}
\usepackage[tight]{subfigure}
\usepackage{algorithm}
\usepackage{verbatim}
\usepackage[noend]{algorithmic}
\usepackage{caption}
\usepackage{cancel}
\usepackage{xcolor}
\usepackage{fancyvrb}
\usepackage{multirow}
\usepackage{hyperref}

\DeclareMathAlphabet{\mathpzc}{OT1}{pzc}{m}{it}

\newcommand{\bitem}{\begin{itemize}}
\newcommand{\eitem}{\end{itemize}}

\newcommand{\beq}{\begin{equation}}
\newcommand{\eeq}{\end{equation}}
\newcommand{\sign}{\text{sgn}}

\newcommand{\ajcomment}[1]{}

\makeatletter
\newcommand{\labitem}[2]{%
\def\@itemlabel{\text{#1}}
\item
\def\@currentlabel{#1}\label{#2}}

\usepackage{bibentry}
\newcommand{\ignore}[1]{}
\newcommand{\nobibentry}[1]{{\let\nocite\ignore\bibentry{#1}}}

\makeatother

\addtocontents{toc}{\protect\setcounter{tocdepth}{2}}

\title{Data Science at the Singularity\\
\small{Version 1.00}}

\author{David Donoho
            \footnote{Department of Statistics, Stanford
              University} 
            }

\begin{document}

\maketitle
\begin{abstract}
\noindent
A purported “AI Singularity” has been much in the public eye recently, especially since the release of ChatGPT last November, spawning social media “AI Breakthrough” threads promoting Large Language Model (LLM) achievements.  Alongside this, mass media and US national political attention focused on “AI Doom” narratives hawked by social media influencers; 
some were invited to tell US congresspersons about the coming ``End Times." The European Commission is now announcing initiatives to forestall ``AI Extinction''.   
 
In my opinion, “AI Singularity” is the wrong narrative for what’s happening now; the remarkable recent happenings signal something else entirely. 
 
{\it Something} fundamental to computation-based research really has changed in the last ten years. In certain fields, progress is simply dramatically more rapid than previously. Researchers in affected fields are living through a period of profound transformation, as the fields undergo a transition to frictionless reproducibility (FR).  This transition markedly changes the rate of spread of ideas and practices, affects mindsets,  and erases memories of much that came before.  

The emergence of frictionless reproducibility follows from the maturation of 3 data science principles that came together after decades of work by many technologists and numerous research communities. The mature principles involve data sharing, code sharing, and {competitive challenges}, {\it however} implemented in the particularly strong form of frictionless open services. Empirical Machine Learning (EML) is today’s leading adherent field, and its consequent rapid changes are responsible for the AI progress we see. Still, other fields can and do benefit when they adhere to the same principles.

Many of the rapid changes due to this maturation are misidentified. The advent of FR in EML generates a steady flow of innovations;
this flow stimulates outsider intuitions that there's an emergent superpower {\it somewhere} in AI. This
opens the way for PR to push all sorts of worrying narratives: not only ``AI Extinction'',
but also the supposed monopoly of big tech on AI research. 
The more helpful narrative observes that the superpower of EML is adherence to frictionless reproducibility practices;
that these practices are responsible for the striking and surprising progress in AI that we see everywhere; 
and that these practices can be learned and adhered to by researchers in whatever research field, automatically increasing the
rate of progress in each adherent field.

\end{abstract}

Frictionless Research Exchange. Emergent Superpower. AI Singularity. AI Hegemony. Brutal Scaling Paradigm.

\pagebreak

\tableofcontents

\newpage

\section{Introduction}

The last 10-15 years witnessed a dramatic transformation of information technology, as the smartphone spread across the planet, spawning omnipresent internet connections, and always-on cloud computing capacity. The result was a unified global computing/communications resource now driving nearly every industry and transforming nearly every human sphere of activities.

Information is now flowing to peasants in developing economies in a fashion inconceivable not long ago; this is driving historically unprecedented declines in global poverty and ignorance.

Our collective intellectual life is also transforming rapidly. Scientists today have a completely different set of ideas about what can be learned from data, and how to go about learning it, 
than only ten years ago.

Readers see facets of this mental transformation in action everywhere they look; they know something has happened, and there’s no going back; but they may not yet perceive larger, implicit, emergent phenomena, causing so many dramatic changes, at the same time, in so many fields.

By this point, each researcher/data scientist knows many prominent individual stories of dramatic recent changes in science’s ability to make sense of data. And we all know where today’s ``standard narrative’’ wants to take this. Deep Learning burst into prominence ten years ago with a triumph in the ImageNet image classification challenge (ILSVRC2012)\footnote{\href{https://qz.com/1307091/the-inside-story-of-how-ai-got-good-enough-to-dominate-silicon-valley}{Quartz: ``The Inside Story of [ImageNet]}''}, and, following massive industrial investment\footnote{\href{
https://www.wired.com/2013/03/google-hinton/}{Wired: ``Google hires Hinton''} \; \href{https://www.theinformation.com/articles/to-find-ai-engineers-google-and-facebook-hire-their-professors}{The Information: ``Google Hires Professors''} \; \href{https://9to5google.com/2016/11/23/google-ai-talent-to-lead-machine-learning-team-in-canada/}{Google 9to5: ``Google builds AI lab''}} – unprecedented for the data science field -- computer vision and natural language processing today exhibit routine capabilities that seemed well beyond our collective aspirations of only a decade ago. Nonspecialist medical technologists can highlight anomalies in diagnostic medical imagery\footnote{\url{https://blog.research.google/2021/09/detecting-abnormal-chest-x-rays-using.html}}; while nonspecialist users can today translate, with high accuracy, texts between many pairs of human languages where there exist very few humans who speak both languages, eg. between certain Pacific island and central African languages \footnote{\url{https://translate.google.com/} \;  \href{https://www.technologyreview.com/2017/05/09/151813/new-software-program-translates-thousands-of-rare-languages/}{Technology Review: ``System Translates Rare Languages''}}.

A parallel story is told for Reinforcement Learning. In 2010 a human\footnote{\url{https://www.computer-go.info/h-c/gobet/index.html}} won a public challenge match at Go on that era’s best publicly available single-computer system. This buttressed the then-prevailing view that, computers like Deep Blue might already have dominated chess play, but for the then-foreseeable future, the reputedly more sophisticated game Go would still be hard for computers to master.  And yet, in less than a decade, computer system AlphaGo burst into prominence, beating world champion Lee Sedol\footnote{\url{https://en.wikipedia.org/wiki/Lee_Sedol}}; AlphaGoZero followed with comparable breakthroughs in many games of skill. No one believes today that the best human can reliably beat the best computer system in many standard games of skill. In fact, sudden improvements in a human player’s track record of human-against-human tournament chess competition give rise to accusations of cheating; such improvements may signal occasional covert reliance on computer systems during what should instead be human-against-human tournament play\footnote{\href{https://www.nytimes.com/2012/03/20/science/a-computer-program-to-detect-possible-cheating-in-chess.html}{NY Times: ``To Detect Cheating in Chess, ... a Better Program''}}.

More specific to core science, the CASP protein structure prediction contest also saw performance levels jump markedly, pushed by the entry of new ideas and energy from the field of Machine Learning\footnote{\href{https://www.theguardian.com/science/2023/sep/21/team-behind-ai-program-alphafold-win-lasker-science-prize}{Guardian: ``Alpha Fold wins Lasker Prize''} \;
\href{https://www.scientificamerican.com/article/alphafold-developers-win-3-million-breakthrough-prize-in-life-sciences/}{Scientific American: ``AlphaFold wins Breakthrough Prize''}}.

There is a tendency among researchers to uniquely attribute these changes to the “Deep Learning Revolution” which, I will argue, misses the key emergent phenomenon arising more fundamentally, and which can be more properly attributed to Data Science. Namely, I maintain that if, counterfactually, empirical deep learning hadn’t burst into prominence in 2012 and if, counterfactually, transformers with attention hadn’t burst into prominence in 2017, basic forces of a data science nature could still have produced many striking improvements in how we go about learning from data. The script would be the same, but different stars {\it would} have been found to be cast in the starring roles.\footnote{Obviously, this is my own opinion, and I mean no disrespect to the very real achievements of Deep Learning and Transformers in their `starring roles'. In the same way, I respect film stars who are well cast and give good performances in their roles in Hollywood films, 
and I praise such performances.  For example, I can praise a specific performance by Glenn Close -- without believing that if Glenn Close had not been not available for a given role, another
actress, say  Meryl Streep, might not also have done a very impressive job.},\footnote{Here is my thinking. About a decade ago certain industrial labs, and later essentially the whole research world went all in on the deep learning toolkit,  {\it in the context of} the data science practices  discussed in this paper.  The investment was enormous: thousands of hires at very high salaries, billions of payroll dollars, and billions of compute dollars. If our civilization invested equivalent resources and offered similar rewards and incentives behind other machine learning strategies, those would also have evolved to improve over time in similar fashion. In fact, the Challenge Problems Paradigm that we discuss farther below has repeatedly, across several decades, driven transitions from not-very-good-performance to human-level performance, across numerous different data types (facial, fingerprint, iris, voice, text) and across different underlying pre-deep-learning technologies. The impressive track record of improving performance in challenges by deep learning throughout the last decade merely repeats a record of improvement we have seen in earlier decades \-- even when deep learning was not in sight. In addition, other strategies exist, far out of today's spotlight, which might have been cast in the same role as deep learning, at the same foundational moment. The `origin story' of the massive shift to deep learning was the performance bump offered by AlexNet in ILSVRC2012. Yet other architectures, more mathematically coherent and explainable, and dramatically more computationally effective, could \-- we now know \-- have delivered the same performance bump, as shown in \cite{zarka2020deep}. 
If --  counterfactually -- such an alternative had been entered in the 2012 challenge, 
it may well have spawned generations of improved descendants through 
a similar wave of investment. Instead, 
as actually happened, AlexNet got the win in ILSVRC2012, and the attention 
went to its underlying technology, which spawned many descendants,
producing the standard narrative of the last decade.}


There is a tendency among the media and lay public to view these many signs of progress
 as signs of an `AI Singularity', where {\bf mysterious superpowers}
 have somehow {\bf emerged}, but are secreted away in hegemon labs, 
 and may soon threaten us with  `AI Doom'. Politicians on the global
 stage are now worried. 

In my opinion, “AI Singularity” is also a misleading narrative, 
 which can’t explain a wide range of new developments and which hides from view the true driving forces.
I will argue that {\it something} fundamental to computation-driven research really {\it has} 
changed in the last ten years, driven by 3 principles of Data Science, which, after longstanding partial efforts, 
are finally available in mature form for daily practice, as frictionless open services offering {\bf data sharing}, {\bf code sharing}, and {\bf competitive challenges}.  

Researchers in fields adhering to these principles find themselves living through a period of profound, rapid transformation. Research life after adopting these principles operates at a faster pace: good ideas spread and are adopted, and improved upon, seemingly with very little friction (eg human labor) compared to earlier decades. We are entering an era of frictionless research exchange, in which research algorithmically builds on the digital artifacts created by earlier research, and any good ideas that are found get spread rapidly, everywhere.
The {\bf collective behavior induced by frictionless research exchange}  is the {\bf emergent superpower} driving many events that are so striking today.

To discuss these data science principles and their singular effects, I first discuss the larger civilizational changes enabling them; I then review the recent public discussion of singularity for eventual comparison with my thesis; and then discuss the dramatic acceleration in research progress which happens when a research discipline adopts the new practices and crosses this singularity.

I then mention actions individual readers can take in response to the arguments presented here, including in research and teaching; and I make predictions about future developments on the other side of the singularity.

\section{The Broader Context}

Our discussion starts with two bits of  background.   

\subsection{Explosion of accessible compute power}

Everyone knows that over the last ten-fifteen years, smartphones have proliferated globally; today purportedly 80\%+ of all adult humans have access to this technology either directly or through friends and associates. Smartphones rely upon  a global computational and communications infrastructure whose rapid construction has been a major civilizational achievement, with myriad consequences and corollaries. These include undersea fiber, satellite internet, arctic data centers, cloud computing, novel database technologies and a massive reconfigurable global workforce of software developers. The global compute budget is perhaps 10000 times greater today than when the smartphone era began. 
Even a negligible fraction of existing global compute capacity could suddenly allow computational scientists to be ambitious at previously unimaginable scales.

Alongside such smartphone-driven growth, some computational science-driven transformations arose independently and organically; these include a massive speculative investment by hardware hegemons in GPUs for numerical computing,  which paid off reputationally with stock market valuations beyond anyone’s expectations. Combined with novel machine learning models and software stacks for constructing workflows, and a whole industry category -- MLOps -- for specifying and deploying workflows, a further 1000-fold increase in compute power has been delivered by orchestrating new ways to organize computation.

Importantly, all this compute power is now available immediately, for a price, to anyone who will pay. 
Originally built to accomodate human habits of sharing information by text and social media  \-- i.e. smartphone apps \-- 
the global compute infrastructure evolved to provide frictionless open immediate access to digital artifacts and compute resources. This {\bf frictionlessly available
infrastructure} can be adapted for many purposes, including computation-driven research.

\subsection{Kurzweil's Singularity Rhetoric} 

In 2005, computer technologist Ray Kurzweil famously forecast that “The Singularity” would take place around 2030, after a factor $10^{14}$  more compute would become available to humanity, compared to its 2005 level. This forecast was derived by stacking many fascinating speculations, at the top end approaching the spiritual and religious. Kurzweil’s audacity firmly planted the idea of “The Singularity” in the minds of intellectuals, after which Computer Scientist Eliezer Yudkowsky and other adherents of the ``Less Wrong’’ online community focused attention on the problem of “AI Alignment” – the worry that AI will one day surpass us, and then, afterwards, be cruel to us. The most vocal worriers are “AI Doomers”; they fear the singularity will arrive with a hard takeoff that will overwhelm us with destructive consequences before we even understand what is happening. 

The  “AI Singularity” interpretation of today’s happenings is traceable to the fact that Kurzweil and Yudkowsky prepared the way by seeding discussion in the public sphere across many years. While many intellectuals might privately be critical, there has not been equally extensive dissemination of counter-narratives. Such counter-narratives, had they been available, might have pointed to the vagueness of the purported `links’ chaining together to produce events leading to death and destruction. Even if the counter-narratives were available, they wouldn’t have the same compelling nature as the pro-singularity narratives, or gain the same traction in mass and social media – explaining why they wouldn’t be produced and developed {\it en masse}, even though the counter-narratives, in my view, would have been correct.

\section{Data Science Matures} 

Data-driven computational research is currently entering a new era, crystallized from the 
maturation of three long-ongoing sociotechnical initiatives spreading throughout science. 

The three initiatives have been aspirations of many data scientists across decades, and the coming-to-fruition moment, where everything needed is now in place and immediately accessible, has been reached by the hard work of many technologists.  Many contributors to this movement had little to do with the field today called “Artificial Intelligence”. There were, of course, quite a few computer scientists involved in these developments and even some who are today considered major figures in AI.


The three initiatives are related but separate; and all three have to come together, and in a particularly strong way, to provide the conditions for the new era. Here they are:

\begin{description}
\item[{[FR-1: Data]}] datafication of everything, with a culture of research data sharing.   One can now find datasets publicly available online on a bewildering variety of topics, from chest x-rays to cosmic microwave background measurements to uber routes to geospatial crop identifications. 

\item[{[FR-2: Re-execution]}] research code sharing including the ability to exactly re-execute the same complete workflow by different researchers.

\item[{[FR-3: Challenges]}] adopting challenge problems as a new paradigm powering scientific research. The paradigm includes: a shared public dataset, a prescribed and quantified task performance metric, a set of enrolled competitors seeking to outperform each other on the task, and a public leaderboard. Thousands of such challenges with millions of entries have now taken place, across many fields.

\end{description}

Each initiative addresses aspirations in scientific research as old as data-driven science.  Researchers have always wanted to access the same data used by their predecessors in earlier studies; or apply the same algorithms from those studies; or to score performance of new algorithms in later studies according to the same metrics used in earlier studies. However, those wishes were generally frustrated in various ways in earlier generations.

In the past decade, all of these initiatives, in their best implementations, became {\bf  frictionless open services}, essentially offering immediate, permissionless, complete, access to each relevant digital artifact, programmatically from a single line of code. Not all friction is yet gone, but the trend is clear: what frictions still remain will vanish soon enough.
  
Today’s researchers will know many specific examples of initiatives  [FR-1]+[FR-2]+[FR-3] coming together 
as frictionless open services in the last decade. Notable ones include:

\begin{description}
 \item[{[FR-1]}] Research Data repositories: Open Science Collaborative\footnote{\url{https://osf.io/}}, 
 Nightingale Open Science \footnote{\url{https://www.ngsci.org/}}; 
 \item[{[FR-2]}] Research code re-execution environments like Jupyter \footnote{\url{http://jupyter.org}}; and 
\item[{[FR-3]}]  Challenge problem platforms such as CodaLab Competitions \footnote{\url{https://codalab.lisn.upsaclay.fr/}}.
 \end{description}
(Prominent commercial examples exist as well;  eg. for  [FR-3], Kaggle, InnoCentive, Sage Bionetworks; for  [FR-2], Google Colab). 

What were once mostly-frustrated wishes of researchers are today achieved realities in several fields, supported by habits and institutions. 
Most importantly, where the behaviors are not yet dependably present, there is no essential obstacle to turning them into everyday habits, except the interest and diligence of participating researchers.

This is a big change from the recent past! The last few decades had been a period of experimentation and prototyping, trying out various approaches to sharing research data, challenge problems, and research code. Sometimes, in principle, we could reproduce the research of others.
But in various ways there was still clumsiness and friction. 

Then came the breakneck development of information technology driving our post-2007 smartphone era.  The emergence of `The Cloud’ gave us tools for web-accessible database management, and source code management for workgroups which made it possible for information technology organizations to be distributed globally yet work together. 

Building on the new capabilities, developers implemented globally accessible code and model repositories, data repositories, and, eventually, globally visible task metric leaderboards. 

These services and their use patterns are well known to modern data scientists. Research communities can now organize themselves to model their research publication process around the use of these services.
  
Among research communities, Empirical Machine Learning (EML) has gone the farthest,  its members today can use accepted commercial services including Kaggle, GitHub, HuggingFace, and Weights and Biases for storing and accessing these artifacts. 
Quite literally, today’s model for academic EML research requires accessing the various artifacts created under [FR-1]+[FR-2]+[FR-3] and then modifying some of them in some way and then sharing those modified artifacts.  Articles describing such research, which we discuss later below, are simply highly stereotyped press releases advertising the availability of these artifacts.

Step back from the EML context for the moment. The same process is ready for immediate deployment by computational scientists more generally. 

\section{Emergence of Frictionless Reproducibility}

The maturation of data science initiatives  [FR-1: Datafication and Data Sharing], [FR-2: Code Sharing and Re-execution], and [FR-3: Challenge Problems] in their 
current form as essentially frictionless services brings us to a phase transition in computation-driven research, to a new era of  {\it frictionless reproducibility} in data-driven scientific research. 
This new research era features unprecedented rates of advance.  

Indeed, today, a computational scientist working in one of those fields where all three practices are standard, can, with a few mouse clicks, and maybe a little extra typing, access everything needed to reproduce some recently reported experimental result from some other institution: code, data, experimental analysis. The scientist can be running some other team’s original experiment or a modified computational experiment within minutes or even seconds of hearing about it. In many cases, the scientist doesn’t even really need to be manually involved; a predefined script can do all the work, completely automating the data and code access, and running the experiment – possibly with novel modifications or measurements layered on top.

In contrast, progress seems notably slower in a field where only 
two (or one, or none)  of these initiatives have yet come to fruition, 
rather than three.

The most common leave-one-out setting is surely {\bf Reproducible Computational Science} (RCS) 
where we combine  [FR-1: Data Sharing] and  [FR-2: Code Sharing],
without [FR-3]. Here there is a scientific question but no underlying challenge problem being considered; we might simply be doing an exploratory data analysis and reporting what we saw, and giving others access to the data and the analysis scripts. RCS lacks the ability to focus attention of a broad audience on optimizing a performance measure. 

A less common but very famous and important setting is the {\bf Proprietary-Methods Challenge} (PMC), with  [FR-1: Data] and  [FR-3: Challenges] but not widespread code sharing by participant teams, even after the fact. The \$1 Million Netflix Challenge (2009), and many large-dollar prize challenges since, are examples. 

Least common by far are {\bf Bring-Your-Own-Data Challenges} (BYODC), ie. offering  [FR-3: Challenges] -  [FR-2: Code Sharing] without  [FR-1: Data Sharing]. This can happen in clinical medical settings where the research focuses on predicting patient outcomes but the underlying data is private and only a few credentialed researchers ever get to see it, under nondisclosure. Here, the research publications may share task definitions and results and also algorithms, without sharing the data. Clinical medical research overall is a field with about 1 million research publications per year.

Without all three triad legs  [FR-1]+[FR-2]+[FR-3], FR is simply blocked. In two of the above leave-out-one scenarios, this is obvious. 
Without  [FR-1 – Datafication and Data Sharing] we would be missing the data; without  [FR-2 – Code Sharing and Re-execution] 
we would be missing the opportunity to inspect and build upon the workflow.\footnote{Again we insist on programmatic access to digital artifacts.}

Less clear is what we might be missing without  [FR-3 – Challenges].  We would be missing the task definition which formalized a specific research problem and made it an object of study; the competitive element which attracted our attention in the first place; and the performance measurement which crystallized a specific project’s contribution, boiling down an entire research contribution essentially to a single number, which can be reproduced.  The quantification of performance – part of practice  [FR-3] -- makes researchers everywhere interested in reproducing work by others and gives discussion about earlier work clear focus; it enables a community of researchers to care intensely about a single defined performance number, and in discussing how it can be improved.

\begin{table}
\begin{tabular}{| l | c | c |}
\hline
If we only have... & We are blocked, because & Example \\
\hline
[FR-1] + [FR-2] & No defined task  & Exploratory Data Analysis \\
\hline 
\multirow{2}{*}{[FR-1] + [FR-3]}  & \multirow{2}{*}{Can't build on code of others} & Netflix Challenge; \\ 
    && DARPA Biometric Challenges \\
    \hline 
[FR-2] + [FR-3] & No Common Dataset & Human Subjects Clinical Research \\
\hline
\end{tabular}
\caption{Leave-One-outs, and what is blocked}
\label{tbl:LOO}
\end{table}

Of course, not every field works this way, but those that do, commonly benefit from very high velocity of progress.  Frictionless reproducibility spontaneously spawns groups of inspired researchers to a tight loop of iterative experimental modification and improvement, often engaging many scientists. This tight loop per researcher, combined with iterative feedback from cross-researcher competition, results in a rapid series of improvements in the performance metric. The outcome of the process is much higher performance on the task metric.

An emergent property is manifest here. The researchers in an adherent  field,
understanding that they are all subscribed to the FR triad, 
think differently about what they do/should be doing, behave differently 
and develop habits and practices that they wouldn't have outside the FR setting.
We see a {\it new institution} arising spontaneously; let's call it 
a {\bf Frictionless Research Exchange} (FRX). 

FRX is an exchange, because participants
are constantly bringing something (code, data, results), and taking something
(code, data, new ideas), from the exchange;
and various globally visible resources \-- task leaderboards, 
open review referee reports \-- broadcast information to the whole
community about what works and what doesn't.
Of course, this is a very different type
of exchange from those involved in financial markets;
it involves intellectual engagement, not money.
Financial exchanges produce price discovery.
Frictionless Research Exchanges produce {\it community critical review}.

Within an FRX, each researcher expects to find out what others have done,
how those results worked, and then can build directly and explicitly on the 
selected work of others; seeking tweaks and improvements over their work. 
A community of researchers working this way initiates a chain reaction of  
reproductions followed by furious experimental tweaking, with each new ascent up the 
leaderboard exciting responses by others and further tweaks and improvements.

Empirical Machine Learning (EML) is, par excellence, the field which has most nearly adopted the 
FR triad in its research model; which has most completely shaped itself to have the habits
and attitudes that create an FRX;
and has benefited the most from FR practices. 

Many computational experiments in EML follow a very uniform structure; 
understood by everyone in the field, the sites and tools (see eg. CodaLab on the noncommercial side;
Weights and Biases, HuggingFace, Github on the commercial side)  
offer easy immediate uniform access to archives of canonical data, models, performance metrics, 
and all the other paraphernalia needed to begin reproducing results as they are announced, 
or developing modified experiments which build on those announcements.

EML research has delivered the numerous stories of remarkable `AI progress’ I noted earlier: in machine vision, 
natural language processing, gameplay, and even protein structure prediction. 
In my opinion, without the last decade's (near-) Frictionless Reproducibility, 
this progress could never have been delivered on the decadal time scale which we have all just witnessed. 
The rate of progress on each problem would have been much smaller, 
because the number of meaningful contributors would be much smaller, 
and uncertainty about the meaning of individual experimental reports 
would have exerted a significant drag on the energy and enthusiasm of the research community.
But also, without the right habits and engagement 
by researchers in the field, the full FRX experience would not have emerged.

EML, by its embrace of frictionless exchange,
has made a very valuable discovery of new habits and mindsets in
research, which can benefit us all, and has revealed the spontaneous 
emergence of Frictionless Research Exchanges, a 
{\bf research superpower} at the heart of the EML story\footnote{Yet, an unsung achievement. Paradoxically, an unsung superpower.}.

The triad of FR practices is not ``the same thing as what EML is doing'';
The EML community seems not inherently committed to the whole FR Triad; 
some hegemons but not others play nice with these practices. 
Protein structure prediction began using the
challenge problem paradigm with the CASP challenges 3 decades ago,
and benefited as we all have from the increasing tendency towards data sharing and code
sharing. 

Other fields might follow these practices and benefit correspondingly.
Scientific fields recently implementing challenges include
NMR Structure Determination \cite{CASD-NMR}, and
Computational Biology \cite{CompBioChallenges}. Recent large-scale funding
for new challenges include an initiative in High Energy Physics \footnote{\href{https://cs.lbl.gov/news-media/news/2022/new-fair-universe-project-aims-to-build-supercomputer-scale-ai-benchmarks-for-hep-and-beyond/}{``DOE awards \$6.4 Million for AI benchmarks for HEP''}} and in
Applied Mathematics (Dynamical Systems)\footnote{\url{https://dynamicsai.org/}, \href{https://www.washington.edu/news/2021/07/29/uw-to-lead-new-nsf-institute-for-using-artificial-intelligence-to-understand-dynamic-systems/}{University of Washington: ``... new NSF institute ...to understand dynamic systems.''}}. 

For three more extended examples, see Section \ref{sec:FREverywhere}.

\section{A Leap Beyond}

In my view, FR and FRX mark a leap well beyond their
three well-known predecessors.

\subsection{ {Beyond} In-Principle Reproducbility.}
Traditionally, computational scientists sometimes offered In-Principle Reproducibility (IPR) in their publications, sharing only some code and data publicly, with additional details embedded in the paper, so that after careful reading and collating of information from various original sources, a researcher could eventually roughly reproduce another’s work. While IPR is hardly frictionless, it stood for some time as a worthy standard to emulate, deserving of approval where achieved.

In the experience of researchers who lived under both regimes, IPR contrasts dramatically with FR.  The common experience under IPR was that during the effort to reproduce, there would be: (a) discovery of many undocumented steps that might offer ‘stoppers’; and (b) many places for previously undisclosed human input and decision where there could be misunderstanding or miscommunication. Consequently, the prospect of success in reproducing reported results was forever in doubt, in turn spawning near-universal skepticism on the part of experienced researchers of results reported by others (the well-known “Not Invented Here” attitude). FR, by removing such undisclosed and undocumented components, makes it much, much easier for researchers to verify and therefore build trust in another group's work.

\subsection{ Beyond Open-Source Software.}

Open-Source Software (OSS)  has been a major presence in the world of information technology for
30 years\footnote{\url{https://en.wikipedia.org/wiki/Open-source_software}}. In its best implementation, which has been widely practiced for decades,
one indeed obtains OSS through frictionless services.
OSS is thus an enabler of FR, an important step in the right direction. 
It enables software developers to build their work on the coding of others; 
this is a crucial part of the usual path to FR, but {\it only one part}. 
Re-executability of a published computational experiment requires {\bf much more} than open access 
to some key software components that were used in that experiment. 
Reproducibility of scientific computations generally requires the exact reconstruction
of an entire computational environment, an entire orchestrated workflow. An important corollary of
the  cloud information technology era, has been the ubiquitous use of virtualization and
containerization\footnote{\url{https://en.wikipedia.org/wiki/Virtualization}}, enabling specific machine instances, and even specific virtual cluster architectures. 
These technologies matured long after the OSS
movement.

In my mind, the Open-Source movement takes us beyond
In-Principle Reproducibility but not yet to full Frictionless Reproducibility,
and also not to the emergence of a Frictionless Research Exchange.
For example, a researcher absorbed in solitary 
pursuits can share code and data, it can even be available frictionlessly,
but without engagement
by a community of other researchers that cares deeply 
about assimilating the best contributions, building upon them, and 
outdistancing them, we will not see emergence of rapid progress.

\subsection{ Beyond Competitive Challenges.}

Challenges alone, without {\it each} of [FR-1: Data Sharing] 
and [FR-2: Code Sharing] {\it and consequent research community engagement},
also can't produce the advantages we are discussing. The typical missing ingredient is
[FR-2], reducing us to the Proprietary-Methods Challenge. 
In such circumstances, we miss the sharing and cross-fertilization of knowledge,
we don't as a community know what has been done to get a method which bests the competitors,
and we don't get the chance for {\it critical assessment} of the methods. It is possible
that a `method' is just a set of random tweaks with no rhyme or reason,
yet can win a particular competition at a particular moment
essentially due to random noise -- some competitions are just that close.
Competitions don't, inevitably, produce lasting results or knowledge.
{\it Community review is what ensures that important advances are recognized and 
propagate.}

Certain communities already take `Open Source' as a given, meaning
data sharing and code sharing.  In such situations, 
[FR-3: Challenges] seem to be the only missing ingredient.
It is then natural to think `Oh, it's all challenges'.
Fair enough. But if that psychology takes hold,
by a `slippery slope effect', we may forget to
secure the assumed givens (Data, Code, Community) and fail
to produce critical assessments.

\section{FR Everywhere!}
\label{sec:FREverywhere}

Crucially, other research fields can benefit from adoption of FR practices and institutions just as well as EML -- or maybe more. 
The FR triad can be, and is being, practiced outside the empirical machine learning world. 

To see this, we do pattern recognition, noting analogies where certain elements in the triad are, 
although different in detail, {\it structurally} identical.

For example, the notion of challenge could change. Here the data and code ingredients [FR-1]+[FR-2] might be as before, but the Challenge task being solved per [FR-3] might no longer be what EML calls a prediction task. 
Mark Liberman, a Natural Language Processing expert at University of Pennsylvania, has been using {\bf Common Task} challenge
to label challenges [FR-3] more neutrally, without implying that they are prediction challenges\footnote{\url{https://www.simonsfoundation.org/event/reproducible-research-and-the-common-task-method/}}.
 
Consider the field of Optimization. There are many algorithms for solving well-posed optimization task instances such as linear programming and quadratic programming, and traditionally one developed personal preferences for one algorithm or another based on intellectual perspectives and mathematical properties. What matters is the speed of an algorithm and the accuracy of the solution. The data science approach in this context would:  [FR-1] share standard datasets which will be used in defining specific optimization problems, [FR-2] share code defining reference implementations, and [FR-3] define metrics (running time, solution accuracy) and maintain leaderboards. Benchopt\footnote{\url{https://benchopt.github.io}} does exactly this, focusing particularly on the algorithms of interest in machine learning; but instead of asking for the prediction performance, as in much EML research, it asks for timing or solution accuracy.  French research funding agencies supported Benchopt with personnel funding and compute access.

Consider an example from Statistical Methodology. Stephen Ruberg and co-authors \cite{Ruberg2022} conducted a challenge to promote development of `subgroup identification methodologies’ ultimately intended for the analysis of data in pharmacological clinical trials. Here {\it synthetic} {\it data} are used, with ground truth known. The authors construct many specified generative models, whose details are (let’s say) sequestered from view, then sample data from those models. Task performance metrics included the ability to correctly diagnose the true underlying groups, because the underlying ground truth synthetic model was known to the organizers, the metrics couldn’t be defined in the usual EML framework. The challenge itself was hosted on the InnoCentive platform, and offered a prize of \$30,000 for first place and \$15,000 for second place. More than 100 competitors vied to predict properties of the generative model. This work and its publication were supported by Eli Lilly.

In physical chemistry, Adam Schuyler and co-authors \cite{NUScon} conducted the NUScon challenge, seeking
reconstruction methods for non-uniformly sampled NMR Spectroscopy.  NUScon was hosted using the NMRbox 
methods platform \cite{maciejewski2017NMRbox}, which allows the creation of workflows using a large number
of different software tools developed by the research community over the years, as well as extensive `workflow glue'
that allows chaining together such methods and tweaking their operation. Synthetic data were created using NMRbox 
based on underlying physical models. Competitors used
NMRbox to code their entry workflows, and administrators used NMRbox and the NMRbox 
compute cluster to implement the contest evaluation. 
The platform provides an (almost) fully-automated workflow for running 
contestant reconstruction recipes on contest data and evaluating solutions.
Several thousand dollars in cash prizes were awarded 
in a NUScon prize ceremony at the annual meeting of the Experimental Nuclear Magnetic Resonance 
conference (ENC). The contest produced community engagement with, and adoption of, methods for nonuniform sampling,
which allow very substantial speedups in data acquisition for high-dimensional NMR experiments. It documented
in a fair and objective way the measured performance of a variety of methods on a common platform.
In addition, NUScon improved awareness of, and adoption of, the NMRbox platform,  and thereby created a
more capable and productive community of researchers.

In each of these three examples, the problems being attacked are not one-one mappings of standard ML problems; but some variant of common task framework can be defined and implemented.

\section{The Roots in Data Science}

The maturation of [FR-1]+[FR-2]+[FR-3] and emergence of FRX did not spring out of a vaccum. Nor out of slidedecks presented to Silicon Valley VC’s who funded Github, Kaggle, and Hugging Face. Nor out of today’s hegemon research labs. Rather, they developed organically from efforts by data scientists and technologists across at least 4 decades, witnessed by my own eyes.

 \subsection{Reproducible Computational Science}

RCS has been developing steadily for decades\footnote{I offer myself as a witness who has seen  and participated in data-driven computing 
first-hand for almost 50 years, including through multiple stints in industry.}.
A few examples \cite{doi:10.1146/annurev-publhealth-012420-105110}:

\bitem
\item  Geophysicist Jon Claerbout began advocating and practicing\footnote{We find the following at \url{https://sepwww.stanford.edu/sep/jon/reproducible.html}: ``In 1990 I set ...  a goal of reproducible documents and results. The basic idea we had is that anyone should be able to reproduce our research results with the software on our CD-ROM ... Many people thought we were reaching too high. I recall some thought we were braggarts, liars, or crazy. Reality showed us it wasn't as easy as it seemed.'' }
 Reproducible Computational Research in the early 1990’s, changing the standards in his field and changing publication practices to enable provision of digital research artifacts.

\item Over the last 20 years, Computational Biologists have been increasingly sharing code and data, 
including genome sequences\footnote{\url{https://www.ncbi.nlm.nih.gov/genbank/}} \cite{GenBank2013}, gene expression data
\footnote{\url{https://www.informatics.jax.org/mgihome/GXD/aboutGXD.shtml}} \cite{GXD2000}, phenotypic data\footnote{\url{https://www.ncbi.nlm.nih.gov/gap/}}, and algorithm libraries like BioConductor \footnote{\url{https://www.bioconductor.org/}}, and workflow frameworks  \cite{GRUNING2018631}.  

\item In the last 10 or 15 years, numerous professional scientific journals and societies have acknowledged the need for, and developed accommodations for, reproducibility of computations \cite{10.1093/biostatistics/kxp014}. Cross-science surveys of this activity are available from reports of the National Academies of Science, Engineering and Medicine; see for example the recent 2019 consensus report {\it Reproducibility and Replicability in Science.}\footnote{\href{https://nap.nationalacademies.org/catalog/25303/reproducibility-and-replicability-in-science}{NASEM: ``Reproducibility and Replicability in Science''}}
\eitem 

The intent has consistently been to share data and code using the technological capabilities and scientific
awareness of the day. As those capabilities advanced, the comprehensiveness of the reproducibility
and immediacy of the access also advanced. In my view, today's RCS capabilities, though they would of course
not been  predicted in detail
by scientists of two or three decades ago, would have been clearly 
anticipated at a high level. A hypothetical scientist of those times might even, had they somehow
gotten a query from the future, have thought
`of course' RCS will come together by 2023!
(I myself would have thought so, even 
had I been asked as long ago as 1995).

Actually, though, many many scientists and technologists had to 
march in the direction of RCS in order to create the culture,
customs, and tools we have today. Moreover, RCS still doesn't happen in every
case. Incredibly to me, advocacy of RCS is spotty in certain fields.

Advocacy of RCS, following RCS practices, and implementation 
of RCS therefore deserve our praise. I won't enumerate
all those I consider instrumental. The already-mentioned report
{\it Reproducibility and Replicability in Science} \cite{national2019reproducibility}
presents an extensive cross-section sampled across
many scientific fields.

Still, for this article, there's one person I'd like to 
spotlight as a standout advocate and implementer.
\begin{quotation}
 {\bf Yann LeCun:}  
Turing Prize winner; Chief AI Scientist of Meta;  Silver Professor at NYU;
and important
thought leader in today's AI information space.  
Among technical leaders in the trillion dollar market cap group,  
LeCun, to me, stands alone as a highly vocal and consistent advocate of 
code and data sharing.  Over the last decade, Meta produced, shared, and supported 
fundamental tools like PyTorch\footnote{\url{https://pytorch.org/} \; \href{https://arstechnica.com/information-technology/2022/09/meta-spins-off-pytorch-foundation-to-make-ai-framework-vendor-neutral/}{Ars Technica: ``Meta Spins off Pytorch''}} and Llama \footnote{\href{https://www.wired.com/story/metas-open-source-llama-upsets-the-ai-horse-race/}{Wired: ``Open-Source Llama upsets AI Horse Race''}}, which are integral to reproducible
computational science in many fields today.
\end{quotation}

As the last touches were being placed on this manuscript,
LeCun spoke to the US Senate Committee on Intelligence,
which wanted to hear about AI and preserving American
Innovation \footnote{\href{https://www.intelligence.senate.gov/hearings/open-hearing-advancing-intelligence-era-artificial-intelligence-addressing-national}{Senate Hearing, Sept. 19 2023}}. As one can imagine, many politicians and 
corporate leaders would be talking in
such a setting about placing technology 
under lock-and-key. Instead (my observation in italics):
\begin{quotation}
{\it LeCun chose to use his precious moments 
of testimony to emphasize the importance of Code and Data Sharing.
He explicitly mentioned these factors as {\bf crucial to the recent rapid advances
in AI.}}
\end{quotation}

In this paper's language, LeCun is implying
that our secret sauce is the frictionlessness
of building on the work of others. This is our 
economic and intellectual engine.

I quibble only in whether RCS alone -- [FR-1]+[FR-2] -- is enough
to get all the rapid advances.
Against my inclinations, I came to believe
one extra factor is essential - [FR-3: Challenges] -- to produce full FR,
and after that, we also need the engagement of a whole
community of researchers with a new type of 
institution -- the FRX -- for emergence of
all the rapid progress we are seeing.
 
\subsection{The Challenge Paradigm }

From my perspective, the {\sl I-didn’t-already-see-this-coming-40-years-ago} ingredient in the FR triad has been the impact of [FR-3: Challenges].  For
at least 50 years it was widely understood that collecting data, and sharing it, could drive data analysis challenges.\footnote{A dataset I manually gathered drove such a challenge problem at the Joint Statistical Meetings in 1983; I have seen challenges with my own eyes for 40 years.}  However, the view of data analysis `challenge’ in early years would have been extremely broad and I would say `humanistic’, and could tolerate a wide range of challenge deliverables, include the challenge of `understanding’ what’s in the data, or the challenge of `revealing an interesting view of the dataset’, or even the challenge of `redefining our ideas of what can be done with the dataset’. A popular term at the time was `exploratory data analysis’, which well evokes the possibility of fun and discovery that inspired many. From this vantage point, what came next was a real surprise.

A very different {\it engineering} view of challenges became formalized\footnote{Ben Recht and Moritz Hardt \cite{hardt2022patterns} have pointed out
that Bill Highleyman, already in the late 1950's, had gathered data and shared it with others and that some of those he shared with
were implicitly competing in a challenge, even though the notion hadn't yet been formalized: \url{http://www.argmin.net/2021/10/20/highleyman/}} in the mid-1980’s for research in natural language processing and biometrics. Researchers were asked to submit entries scored by specified, predefined metrics. Leaderboards were instituted and winners declared algorithmically. This approach to challenges was radical in its clarity and simplicity, but also in its narrowness of focus.\footnote{I recommend Mark Liberman's presentation\\ \url{https://www.simonsfoundation.org/event/reproducible-research-and-the-common-task-method/}}

In a series of projects, many funded ultimately by the Pentagon (DARPA) in the 1980’s through the 1990’s and beyond, 
challenges were mounted in speech processing, biometric recognition, facial recognition, and other fields. 
Funding,  often to NIST \footnote{\url{https://www.nist.gov/biometrics}}, covered data collection and curation, and challenge contest administration. 
Sometimes, winners of challenges were selected for follow-on DARPA grants and research contracts. 

In a major departure from the principles advocated in this article, some DARPA challenges followed the Proprietary Challenge model, which allowed contestants to keep their code and working methods confidential, essentially meaning that contestants learn leaderboard results, but not necessarily about the winning models or fitting procedures. Such an implementation of [FR-3: Challenges] is incompatible with  [FR-2: Code Sharing], and hence with frictionless reproducibility, showing it makes a difference when all three of [FR-1]+[FR-2]+[FR-3] are specifically present. This longstanding tension is important at the present moment in empirical machine learning, where some internet hegemons take a proprietary line.

Other data science fields instituted challenges organically without Pentagon instigation.  The CASP protein structure prediction problem is the longest-lasting scientific recurring challenge problem venue. The sequence of a certain protein is known; predict its 3D structure.  This competition has been around roughly as long as the world-wide-web and web browsers; its website today at \url{https://predictioncenter.org/} documents the (peri-) biennial competitions since 1994 and the evolution of predictive solutions.  Many outside-of-proteins data scientists’ first inkling of this competition followed a recent episode of hegemon engagement; DeepMind entered into one of the CASP competitions, deployed overwhelming computational resources, got their win, and issued their corporate PR, to the applause of the capital markets\footnote{\href{https://venturebeat.com/ai/deepminds-alphafold-wins-casp13-protein-folding-competition/}{Venture Beat: ``AlphaFold wins CASP''}}. But the competition has been ongoing for 3 decades, following [FR-1]+[FR-2]+[FR-3] rigorously, and gradually improving protein folding; without a corporate PR budget then or now.

Many data scientists ought to be mentioned as heros of the challenge paradigm, across science and technology, and across decades. 
I list 3 important figures, in last-name-alphabetical order.

\bitem

\item {\bf Isabelle Guyon} of Université de Paris--Saclay \footnote{\url{https://guyon.chalearn.org/}}, 
who collaborated on the original MNIST dataset of the late 80’s and early 90’s and co-authored 
a foundational paper on computer handwritten digit recognition \cite{bottou1994comparison}.  
Over the last 20 years she has been heavily invested in many prediction challenges
\cite{Guyon2006,guyon2008analysis,guyon2011unsupervised,guyon2015design,liu2021winning,jiang2020neurips}.
 A recent example is the 2Million Euro Horizon prize for big data technologies\footnote{\href{https://research-and-innovation.ec.europa.eu/funding/funding-opportunities/prizes/horizon-prizes/big-data-technologies_en}{Horizon Prize}}.

\item {\bf John Moult} of the University of Maryland, a molecular biophysicist who has been a sparkplug of the CASP protein structure 
prediction contests since the very beginning, i.e. the early 1990's
\cite{moult1995large,kryshtafovych2010casp,moult2005decade,moult2006rigorous,kryshtafovych2014casp10,moult2018critical,kryshtafovych2021critical}.

\item {\bf Jonathon Phillips} of NIST, who played a crucial role on many of the pioneering ‘90s and 00’s 
NIST/DARPA challenges in vision and biometrics, including data collection and contest organization
\cite{Phillips1998feret,phillips2000feret,phillips2002gait,phillips2005overview,phillips2009overview},
with challenge involvement continuing up to the present time.

\eitem

Although they have very different professional backgrounds, they have in common an
exceptional clarity on the importance of challenges
and exceptional persistence in promoting this paradigm in their domains.

From a later generation, I will also mention:

\bitem
\item {\bf Percy Liang} of Stanford; he had the farsighted idea in the 2000’s to create a sort of 
`operating system for conducting challenge problems’ that is today’s CodaLab
\footnote{\url{https://worksheets.codalab.org/}}, and which administers a large number of academic challenges annually
\footnote{\url{https://codalab.lisn.upsaclay.fr/}}. 
\eitem

These data scientists, and many others I don’t have space to mention, have shown how to make data science a {\bf science, full stop}. 
Their vision is that data science research projects making full use of [FR-1]+[FR-2]+[FR-3] 
are truly empirical science; while research projects without all three will fall short in some way.  
In fact, Isabelle Guyon, in her 2022 keynote
address at NeurIPS \footnote{\href{https://slideslive.com/38951380/keynote-talk-by-isabelle-guyon-and-evelyne-viegas-ai-competitions-and-the-science-behind-contests?ref=speaker-23300-latest}{Isabelle Guyon: NeurIPS Keynote}}, has chosen to emphasize exactly this: through challenges, 
Machine Learning has become an empirical science.\footnote{Many mathematical scientists ask me: what is {\it conceptually} new in EML, 
really outdistancing the various contributions that mathematical scientists put forward long ago? 
In my view it is easiest to explain this as the consequences of a community embracing the FR triad and
 reaping the consequences.}

So the social practice, of challenges propelling research, wasn’t born yesterday; it has been developing and refined across four decades. Pioneers saw early on, and soon made others understand, that this would be a transformative development.

\subsection{Surprising Reactions}

In a sign of its potency, [FR-3: Challenges] elicits a range of surprising reactions.

\subsubsection{Reaction 1: Wild Enthusiasm}

Mark Liberman has made clear that already in some of the earliest contests, the challenge paradigm tapped into a very special, previously unexpressed human energy. Leaderboards game-ified research and exerted a deep grip on the moment-by-moment attention of many participants of early challenges. They eagerly dropped into a tight loop of tweak model and data, submit new answers, try again. It was understood that performance improvements were happening at a much faster pace, even in the humble early days of this paradigm. It was also clear from this early experience that, if data and code sharing grew, and leaderboards went global, this human energy would increase in staggering ways.  Exactly as happened in EML over the last decade.

\subsubsection{Reaction 2: Shock and Disbelief}

Quite a few mathematical scientists (including myself) have found it initially difficult
to accept that a great deal of work in the empirical machine learning community
was not supported by careful formal analysis and derivations
e.g.  theorems from mathematics or from CS-theory.
Many math scientists (also including me, intially) had difficulties understanding  
that a whole scientific community could be
based entirely on the foundations of
reproducible computations
and challenge leaderboards.\footnote{i.e. literally an epistemology based on RCS=[FR-1]+[FR-2] as well as Challenges [FR-3].
See Section  \ref{sec-Epistemology} below.
Many also initially miss  the emergent superpower,
the  community behavior that sets in as FR becomes FRX.
} 

In processing my own sense of dislocation, I turned
to the famous article of  Alon Halevy, Peter Norvig,
and Fernando Pereira {\it The Unreasonable Effectiveness
of Data} (2009) \cite{halevy2009unreasonable}. This squarely took aim at Eugene Wigner's
{\it The Unreasonable Effectiveness
of Mathematics in the Natural Sciences}  (1960) \cite{Wigner1960},
an article of some veneration among those trained
in the mathematical sciences over the last 60 years.
Halevy et al.'s article is best viewed as a 
provocation rather than an attempt to convince doubters,
but it does
brutally explain the new rules of the game.\footnote{Matan Gavish
and I are preparing a lengthy discussion about the contrast
between Digital and Mathematical worldviews that aims to be 
more instructive, and less merely provocative, than Halevy et al..}

\section{Acceleration towards Singularity}

The argument so far is an argument that [FR-1]+[FR-2]+[FR-3] have recently come together, 
after lengthy gestation in plain sight, and this combination \-- 
in the presence of the right research community practices \--
unleashes an acceleration of computational-scientific research. 
But acceleration is everywhere, not just in research; 
for inhabitants swimming through the 2020’s decade, it surrounds us like water. 

\subsection{Acceleration Everywhere}

Innovations of all sorts are today adopted more easily and spread more rapidly than ever. The spread of memes on the internet was recognized as a new phenomenon already in the 1990’s; the sudden virality of slogans, posters, or videos that began to be seen at that time is today a dependable feature of the information space; each day on social media, the `trending now’ feature presents that day’s viral topics, likely to be replaced tomorrow by other viral topics, often unpredictably.

At the heart of viral meme spread is the `single-click transmission’ of ideas. Some saying or image is appealing, and the ease of transmitting that message to others is so effortless, just a few clicks; the viewer overcomes inertia, and forwards it to others; initiating or continuing a chain. This is then repeated, in some cases almost endlessly.

Some observers see a biological drive at work; it is common to say that each participant in the chain gets a `dopamine hit’ from her/his `discovery’ of the content in their in-box, followed by a fulfilling `action’ of reposting or forwarding\footnote{\href{https://sitn.hms.harvard.edu/flash/2018/dopamine-smartphones-battle-time/}{Harvard Science in the News: ``Dopamine and Smartphones''}}. Some participants speak darkly of their `addiction’ to participating in such chains as caused by `dopamine craving’ \cite{lustig2018hacking}. Others praise, and seek out, the `flow state’ induced by participating in the single-click transmission of ideas. 

We are now so used to frictionless spread of sayings and images, some quite arresting, entertaining, and shocking, that many of us expect a daily parade of amazement.

Mass culture has been practically overwhelmed by this development. Billions of people have smart phones, which expose them to social media for hours a day\footnote{\url{https://www.statista.com/statistics/433871/daily-social-media-usage-worldwide/}}, during which they experience the cresting and breaking of waves of viral memes.

This has upset traditional activities like politics \footnote{\href{https://www.brookings.edu/articles/how-tech-platforms-fuel-u-s-political-polarization-and-what-government-can-do-about-it/}{Brookings Institution: ``Tech and Political Polarization''}}. 
Popular `grass roots’ movements can grow virally and crest beyond the reach of any traditional debate or clash of ideas. 
The attention-grabbing capabilities of such viral waves are breathtaking.  
Originally ``fringe'' ideas can suddenly emerge, seemingly from nowhere,
to capture significant mindshare.

The recent unprecedented dynamism of mass culture is happening all around us, all the time; but we may forget that it is rooted ultimately in the technological possibility for delivering (almost) 
frictionless spread of sayings and images. Today’s frictionless regime is invisible to us fish; it is water. 

\subsection{Friction spoils everything}

In contrast, friction – drag on the effortless sharing of information -- is really, really visible to us. It breaks the flow, stops the dopamine. Today, many have no patience for friction, of any kind.

One can see complaints about this on social media: `receipts or it didn’t happen’, `URL or I don’t care’. Commenters want the experience of accessing digital content with a single click. If there is no digital content to access efficiently, participants are frustrated and share their frustration, vocally; perhaps even with digital tokens of their frustration, such as emojis or GIFs. 

Putting content behind `paywalls’ has the same effect. Some participants will react strongly against it. Some may not want to pay, but many simply don’t want friction!

\subsection{Virality in Data Science}

Frictionless reproducibility of scientific computations is the analog, for research, of the frictionless transmission of internet memes.

FR aligns powerfully and naturally with the habits and practices all of us experience from digital culture at large. 
A new methodological tool published in a way that subscribes to  [FR-1]+[FR-2]+[FR-3] may spread across a field like wildfire.

I have seen this up close in single-cell RNA-Seq data analysis. A paper \cite{Seurat} in a leading journal used a new software package, Seurat, and the data and code were packaged in a very clean way\footnote{\url{http://seurat.r-forge.r-project.org/}}, with a user interface that made it extremely easy to reproduce their results – but also to apply the same methodology to fresh data from another project. The package spread across the RNASeq research landscape rapidly and it soon became de rigueur for visualizing RNASeq data in publications. 

In a series of interactions with the computational biologist I was working with,
the narrative quickly went from ‘what is this new tool? Can I trust it?’ in an initial meeting to, a few meetings later, ‘I was just at a conference, Seurat is everywhere; you now have no choice, you must now present your data using Seurat, or no one will take you seriously’.

After diving into Seurat’s technical ideas, I saw that a key driver of such viral spread must be simply the ease of adoption of the package and the attractiveness of the produced plots, almost divorced from other considerations. The package was very well designed for ‘unboxing’ by newcomers, effortlessly delivered colorful presentations of some supplied datasets, and following the rules for creating general purpose software for use with other datasets. In my opinion, this ease of adapting the tool to other datasets definitely propelled Seurat’s adoption.

By the way, Seurat was not the lone option; many competing methodological tools were available. I studied some of them and came to the conclusion that Seurat offered the least `reproducibility friction’, and so it spread.

\subsection{Onset of Frictionless Reproducibility as a Singularity}

Today’s onset of Frictionless Reproducibility will have major consequences for scientific research. 
To see this, consider lessons from  {\it The Singularity is Near}, \cite{kurzweil2005singularity}. 
Kurzweil quotes one of the 20th century’s most prominent mathematicians,
John von Neumann: 
\begin{quotation}
The history of technology ... gives the appearance of approaching some 
essential singularity in the history of the race, beyond which, human affairs,
as we know them, cannot continue.
\end{quotation}

\begin{figure}[htp]
\centering
\includegraphics[height=3in]{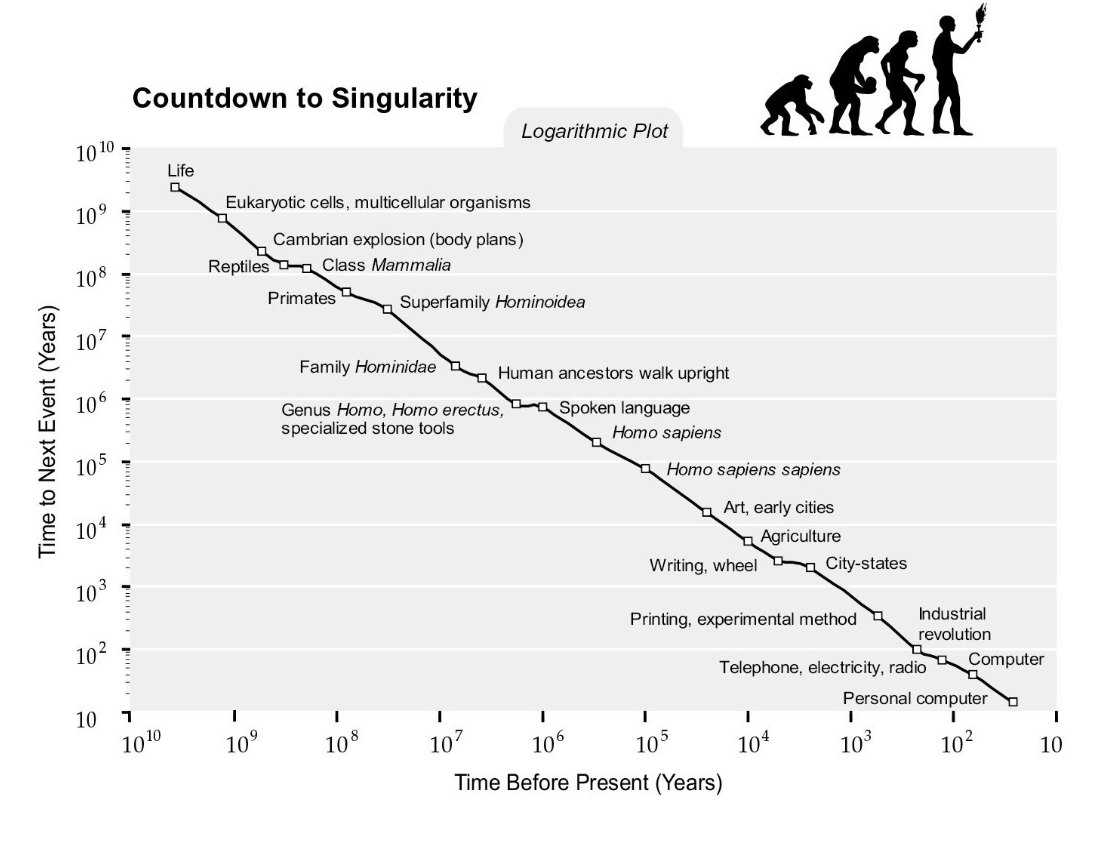}
\caption{from Kurzweil, 2005}
\label{Fig:Countdown}
\end{figure}

Von Neumann introduces the idea that a singularity is coming.\footnote{
 Von Neumann died in 1957. He was presumably at the time of this quote 
gloomy from his own impending early demise and from the 
Soviet Union's apparent willingness to use nuclear weapons
 including H-bombs it to advance its imperial ambitions. Von Neumann
makes two unnecessary leaps: (a) the terminology `essential singularity' \-- borrowed from complex analysis \--
overdramatizes, (b) the concept of `could not continue' \-- also borrowed from complex analysis \--
again goes too far. 
Von Neumann's fears have not yet materialized;
the global diplomatic order has been able to inject {\it friction} into the
development of nuclear weapons, including outright prohibitions on testing, and 
sanctions on nuclear violator countries, such as North Korea.  A more objective and
less dramatic statement would have been: "The history of technology gives the appearance of approaching
... {\it a new era of unprecedented rapid change}." But of course if von Neumann's phrasing had been
less dramatic we would not be discussing it today.
}  
But when?  Kurzweil presents Figure \ref{Fig:Countdown}.

Here Kurzweil’s vertical axis presents “time between notable events” and the horizontal axis 
shows time before the present day. The interval between notable events is dropping towards zero, signaling acceleration in the pace of human affairs.\footnote{Dr. Michael Feldman points to the arbitrary character of the `notable events' in Kurzweil's list, suggesting that
their selection might be biased to prove the rhetorical point being made. } 

Combining the last two exhibits, Kurzweil's rhetoric suggests that: 

\begin{quotation}
\sl when the time between notable events {\bf drops to zero}, we are at {\bf singularity}.
\end{quotation}
The `reaching zero moment' is quite explicit
in both Von Neumann's and Kurzweil's account; it involves observables
we can directly sense. I adopt this as my {\it criterion for `dating a singularity'}.\footnote{This notion of
 singularity differs from another one of Kurzweil's, mentioned earlier. 
That other one points to the moment 
 when the amount of non-biological computation
first exceeds biological computation;  
but this seems not directly observable, while ``time between major innovations
reaches zero'' seems something we can all sense, and hence more empirically valid. In Kurzweil's AI context the 
difference between the two criteria is perhaps a decade.},\footnote{
Kurzweil is completely correct to call our attention
to the trend towards a decreasing time interval between key innovations,
and to attach importance to the fact that 
this time interval is currently approaching a vanishing point.
However, after following along on Kurzweil's voyage this far, I 
jump off his train, and so should you.},\footnote{Also important to my thinking:  singularity 
can have a neutral interpretation in mathematics.
It doesn't necessarily involve the world
 blowing up or ending abruptly.
In Fourier analysis,
a function $f_\alpha(x)$ of a real variable such as
$  \sign(x-s) \cdot  |x-s|^\alpha$ for some fixed $\alpha$
can be said to have a singularity at point $s$.
This includes $\alpha \geq 0$ which does not
`blow up' as $x \mapsto s$.
Singularities don't have to diverge.
There just has to be a noticeable break in behavior,
before versus after. 
(Of course, in complex analysis the term 
singularity does generally involve `something blowing up'.)

Perhaps the reader thinks that `inflection' makes more
sense than `singularity'. And yet. Is the recent rapid change in
machine learning performance, across many tasks and
data modalities, a mere inflection? It seems we saw
a {\it rupture} in performance between the adjacent decades of
the 2000's and 2010's.
}

Compare to Frictionless Reproducibility in computational science. 
FR heralds a drop to zero of the human effort required to reproduce a computational result. This micro-phenomenon (time for a researcher to reproduce one result dropping, essentially to zero) in turn drives a macro-phenomenon -- the time for a field to globally adopt a new dominant methodology -- also dropping, essentially to zero.  As we saw in the rapid embrace of Seurat for Single Cell RNA data analysis.

I see analogies to the physicist’s idea of superconductivity, where an electrical conductor's resistance drops to zero. Many have explained that easy access to superconductivity could one day have amazing consequences for the energy industry and the world economy\footnote{For example: {{\href{https://theconversation.com/room-temperature-superconductors-could-revolutionize-electronics-an-electrical-engineer-explains-the-materials-potential-201849}{Article: ``Room-Temperature Superconductors could Revolutionize Electronics''}}}}. I likewise sense that frictionless reproducibility can have, and already is having, impressive consequences. 

Everywhere we look, progress in science and technology is speeding up. At the same time, we see a transition to frictionless exchange of digital research artifacts. 
At the macro-level, speedup; at the micro-level, frictionlessness.

The mRNA vaccine story of 2020 is a well-known emblem of science acceleration. Early in the COVID-19 pandemic we were told that novel vaccine development has never taken place in less than a decade, and yet in about a year hundreds of millions of people were already vaccinated, in the US mostly with mRNA vaccines \cite{bourla2022moonshot}.  A key part of the story, often left out, is the frictionless spread of information about the SARS-COV-2 virus, most importantly its sequence. The virus’ RNA sequence data were published on virology websites in mid-January 2020 and within days virologists all over the world were analyzing the sequence and very soon had vaccine candidates\footnote{\href{https://www.ncbi.nlm.nih.gov/pmc/articles/PMC7154514/}{Cell Host Microbe: ``Genome Composition of Novel Coronavirus Originating in China''}}. The new mRNA technology fit perfectly with this situation as it allowed virologists to design a vaccine candidate directly from digital sequence – a comparatively frictionless process. The friction in the end-to-end process was all in the traditional political, regulatory, manufacturing, and public outreach realms. The scientific and technological progress was, relatively speaking, instantaneous\footnote{\href{https://www.nature.com/articles/d41586-020-03626-1}{Nature: ``Lightning Fast Covid Vaccines and what it means for other Diseases''}}. 

Key enablers of this rapidity were  [FR-1] data sharing (of the virus’ RNA sequence) and widespread  [FR-2] code sharing (of algorithms that could analyze and translate that sequence in various ways).

We earlier discussed the staggering rate of recent progress in large language models. The sudden emergence of GPT-4 and chatGPT in last year’s public discourse spotlights one company’s work; but it is better viewed as a evidence that the NLP field as a whole has made very dramatic progress across many tasks and challenges in the last decade.  Consistent with our theme, the habits and institutions of  [FR-1]+[FR-2]+[FR-3] are, as we expect, very strong in Natural Language Processing (NLP). The availability of massive language datasets ([FR-1]), the sharing of architectures and fitted models ([FR-2]), and the popularity of challenges ([FR-3]) are heavily present in this field; their synergy and associated institutions, including 
CS conferences\footnote{\url{https://2023.aclweb.org/}} and Open Review of submissions\footnote{\url{https://openreview.net/} \; \url{https://aclrollingreview.org/}}  shaped the community
 of NLP researchers into a Frictionless Research Exchange.

Again, frictionlessness at the micro level, rapid progress at the macro.

Combining the above comments, and applying the criterion given above,
I propose we are at a  singularity \-- the {\bf reproducibility singularity}.
I don't propose dramatic claims of mass unemployment 
and human extinction. I do propose instead fundamental changes
affecting computation-driven research and its rate of progress.

\section{Other Viewpoints and Mindsets}

The narrative so far knits together
many key facts about our new information technology landscape
for computation-based research
and the evolving scientific method.
Before judging this narrative,
let's discuss other narratives and
viewpoints, and compare baked-in assumptions.

\subsection{Counter-Narratives}

Social media influencers harvested clicks and likes over the last year promoting an `AI storyline’, hinting at the hegemon labs’ would-be arrival at the threshold of  the `AI Singularity' 
This meme exploded in November 2022 around the time chatGPT was released, and hegemons 
been the happy recipients of much free media attention.

The storyline provoked plenty of enthusiastic ‘AI Doom’ and `AI Risk’ social media influencers and influencees. 
Key events include an open letter from AI personalities advocating a pause in AI research\footnote{\href{https://www.theguardian.com/technology/2023/may/30/risk-of-extinction-by-ai-should-be-global-priority-say-tech-experts}{Guardian: ``Risk of Extinction by AI ... say experts''. May 30, 2023.}}, and a US Senate hearing on AI regulation\footnote{\href{https://www.nytimes.com/2023/05/16/technology/openai-altman-artificial-intelligence-regulation.html}{Times: ``... Urges AI Regulation''. May 16, 2023.}}, accepting testimony by corporate reps from OpenAI and IBM, and a social media `AI Risk’ personality. The European Commission announced in mid-September:
``Mitigating the risk of extinction from AI should be a global priority.''\footnote{\href{https://www.sundayworld.com/news/world-news/ai-extinction-risk-should-be-a-global-priority-eu-parliament-says/a1172701390.html}{Sunday World: ``AI extinction risk should be a global ‘priority’, EU parliament says.''} September 15, 2023.}
The Telegraph of London, on September 25, 2023, quotes PM Rishi Sunak as saying that `Britain has one year to prevent AI 
running out of control'. Sunak is convening a global roster of hegemon tech leaders in November 2023 in Bletchley Park,
stomping grounds of Alan Turing and the Enigma codebreakers. 

Hegemon PR favors the perception that only hegemon research labs can `do modern AI’ and that {\it they} are the engine of AI progress \footnote{\href{https://www.economist.com/business/2023/03/26/big-tech-and-the-pursuit-of-ai-dominance}{Economist: ``Big Tech ... AI Dominance''. March 26, 2023.}}.  
Hegemons become the focus of the global media  and political class \footnote{\href{https://www.washingtonpost.com/technology/2023/09/13/senate-ai-hearing-musk-zuckerburg-schumer/}{Washington Post: ``Tech Leaders..  Call for Government Action on AI''. September 14, 2023.}}.

There is a kernel of truth in this \cite{abdalla2023elephant}.
Industrial information technology often follows a {\it brutal scaling} paradigm,
simply throwing more and better compute resources at problems 
(some academics also advocate this; see passage about `The Bitter Truth' below).
Aligned with this tendency, hegemon labs’ predictive models 
have been growing in size at a breathtaking pace, by factors of hundreds of thousands or even millions over the last decade.
Today’s LLM model trainings require massive compute budgets (hundreds of millions or even, soon, billions of dollars annually, so that their researchers can train models with trillions of data and parameters). Such scaling-up efforts go far beyond the compute budgets of any academic lab. The resulting perception that `scale is the path forward’ leads to the perception that the internet hegemons `own AI’ and that investors can own some of it, by buying shares in the hegemons.

Adherence to the brutal scaling paradigm also leads to crushing hardware demand and escalating prices for the most state-of-art hardware, driving lengthy GPU order backlogs and  pushing GPU manufacturer Nvidia into the trillion-dollar-stock-market capitalization club.

The `corporate ownership' narrative hides from view the contributions of Data Science and the recent maturation of Frictionless Reproducibility. Data Science, despite its pervasive nature in modern corporate operations, is not a corporate interest, has no market capitalization and no PR agency. It is easy to steal credit from Data Science. 
The triad  [FR-1]+[FR-2]+[FR-3] created the conditions under which current AI technology crystallized and was developed globally, and created a process by which thousands of researchers globally could rapidly access that technology and rapidly be trained in its use and deployment. It created an arena \-- an FRX \-- where unprecedented numbers of qualified researchers cared about open research problems and actually attempted to compete at finding a better model or better training regime. 
Also, in the normal iteration of its working, it drove adhering fields to speed up their progress radically. 
In my opinion, {\it {\bf this} is the emergent {\bf superpower}, the source of the acceleration, being {\bf confused with AI Singularity}}.

Brutal scaling {\it seems} to offer hegemons a way to make an end-run around the FRX superpower, in line with their corporate interests. In a stark throwback to regrettable practices predating [FR-1]+[FR-2]+[FR-3], hegemon labs (except Meta) often don’t share data or models \cancel{[FR-1]}-\cancel{[FR-2]}, so the outside world often can’t reproduce their work. Hegemons are effectively asserting proprietary rights on technology that started life as a public good and reverse the steady and hard-won progress by data scientists toward today’s frictionless reproducibility. They also may place themselves outside the Challenge Problems regime  \cancel{[FR-3]}, benefiting -- strange but true -- because the public discourse about the new LLM’s becomes thereby uncritical. Hegemony becomes more dominant, when the structure and performance of LLM’s is not consistently defined, measured, shared or discussed.  Political leaders and Board Chairpersons spend weekends with chatGPT and come away impressed.  Sentiments rule.

The end-run offered by brutal scaling is likely temporary. A global community of empirical machine learning researchers now exists outside of the proprietary labs, adhering to  [FR-1]+[FR-2]+[FR-3], and conducting open research to narrow the gap with hegemon laboratories. Even in the most ambitious LLM settings, shared datasets are very rapidly coming online and shared models have improved their performance markedly in recent months. 

As one hegemon staffer famously leaked\footnote{\url{https://www.semianalysis.com/p/google-we-have-no-moat-and-neither}} recently, `we have no moat’, meaning, 
in our language:

\begin{quotation}
\sl `[in this new FR world] we [hegemons] have no moat’.  
\end{quotation}

Under FR, each research achievement by one specific lab will very soon be surpassed somewhere, and the achievement will rapidly diffuse everywhere. 

We thus see two competing narratives about recent events: the hegemon-favoring `brutal scale is the one and only path’ and this paper’s `frictionless reproducibility has now emerged, has been responsible for great advances and will take us forward’. 

Brutal scaling-up of computations and datasets has worked in the last ten years, but this was not guaranteed; it was a speculative bet which paid off by amazing luck. For the next decade, the brute force approach is unlikely to pay off as it did recently. The costs of training simply can’t scale by another 100 or thousand, as they did last decade. 

Frictionless reproducibility, on the other hand, is a permanent gift that keeps on giving. Among its benefits is the enlistment of a global community of researchers tweaking and tinkering, improving upon existing research achievements. Again and again, across decades, such enlistment propelled advances in protein folding, in speech, in biometrics,  in natural language processing, and elsewhere. More research fields should follow this model.

As time goes on, I expect the brutal scaling narrative to step back into the ranks, and the new frictionless reproducibility regime, with its respect for and access to human creativity, to become ever more and more evidently the sign of modernity. Hence, this article. 

\subsection{The Exponential and the Logarithm}


Look at a plot of predictive performance on EML Challenge problems over the last decade versus dataset size and/or model number of parameters\footnote{Eg. Figures 3 and 4  of Deepmind's Chinchilla paper  \url{https://arxiv.org/pdf/2203.15556.pdf}}; you will see, first of all, what the hegemons want you to see: staggering increases in dataset and model sizes. You can also see what they would downplay: a breathtakingly bad exchange rate. In classification tasks, the rate at which the large models improve with resource expenditure is either a power law with power close to zero – behavior as bad as I’ve ever seen documented in empirical scientific work – or, even worse, it is actually zero, which is to say logarithmic. 

The first 800-pound gorilla many in AI avoid discussing: if brutal scaling is the only path to better models, we won’t be able to afford it. This is not a normative statement; it is reality. We already use a noticeable fraction of global energy production on global data centers; adding extra zeros to our budget for training expenses in order to get diminishing bumps in performance is simply not in the cards. The old wealthy nations have aging populations who will be cashing in their retirement stocks soon; even new leaders like China will do so soon enough. We won’t spend our retirement savings on tech hegemon mindless pursuit of scaling; we have other uses for accumulated wealth, such as medicine and nursing.

The second 800-pound gorilla:  AI proudly, willfully, {\it has no ideas} and {\it doesn’t want any}.  The blog post\footnote{\url{http://www.incompleteideas.net/IncIdeas/BitterLesson.html}} “The Bitter Lesson” by Rich Sutton explained in 2019 that AI doesn’t want any help from intellectuals and their pet ideas;  it views attachment to ideas with suspicion. Instead AI adherents are hectored only to use the most basic, most general principles (e.g. give me more data, give me bigger models, climb uphill) and apply brutal scaling. In this view, the Moore's law cavalry will always arrive, just in time, to rescue the imperiled scaling paradigm; listening to intellectuals and being swayed by their ideas to get clever computation boosts here and there will only distract AI adherents and delay getting on with the true business at hand: brutal scaling.\footnote{Contra Sutton, we can point to ideas which {\it have} made a difference. The Fast Fourier Transform was an idea about how to apply a specific dense matrix in order $O(n \cdot log(n))$ time rather than $O(n^2)$. It emerged in the 1960’s to broad acclaim as a fundamental advance which became heavily used in data compression, data transmission, cryptography, and signal and image processing. Many of the advances of the information society between 1970 and 2010, including global communications infrastructure and smartphones, would never have happened on such a time scale without this one idea.
But there were many other essential ideas contributing as well; for example, dynamic programming for decoding (Viterbi's algorithm)
was heavily used in communications infrastructure, and converted overwhelming combinatorial search problems into
manageable ones.
 In Sutton’s view, Moore's law can do everything.
But Moore’s law only worked its magic during its lifetime. because in parallel, 
elegant ideas were enabling existing computational technology
to work much more effectively,
enabling new consumer products
with the computing technology of the day,
creating future consumers of what Moore’s law could yield {\it at that time}. 
The growth of consumer society really did 
depend on elegant ideas like the FFT, the Viterbi algorithm, and many others, to achieve its massive development 1970-2010. 
Absent those, there wouldn't have been an eager public buying the fruits of Moore’s law. In short, we wouldn't have had
Moore's law for 40 years if we had only had Moore’s law to rely upon between 1970 and 2010. 
People would stop investing in `more Moore's Law' as soon as the other technology
wasn't there to support it.}

The brutal scaling paradigm {\it sort of} made sense for half a century during which Moore’s Law held sway, and we lived in a scaling paradise with continual improvement of CPU processors and memory density.  But Moore’s Law quit working a decade ago, and the end of GPU scaling is already in sight\footnote{\url{https://epochai.org/blog/predicting-gpu-performance}}; 
yet AI remains attached to this obsolete conception.\footnote{Sutton does acknowledge that Moore’s law is finished, but in a science fiction move worthy of Ray Kurzweil, postulates a would-be generalized Moore’s law that always arrives just in time to keep the AI `we don’t need ideas’ narrative viable. The inconvenient truth is that we don’t have any concrete path to the next scaling miracle. As they say: Hope is not a strategy.}

While the AI Singularity narrative faces an uphill slog in the face of logarithmic returns on investment, 
the FR Singularity narrative has a downhill romp, effortlessly propelled by exponentials in its favor. 

The Frictionless Reproducibility triad [FR-1]+[FR-2]+[FR-3] fosters a global community of methodological research contributors, a global audience for the fruits of their methodological developments, and thereby a global research exchange, an FRX. The exchange identifies good ideas and endows them with positive exponential growth through a global user population. Iterating this, a channel is created through which better and better performance is achieved.\footnote{Julian Simon \cite{Simon+1996} points to the fact that human talent and ingenuity have delivered again and again. Challenges, as part of the full FR/FRX package,  open up research problems to everyone, drawing on Simon’s  `Ultimate resource’. 
\url{https://en.wikipedia.org/wiki/The_Ultimate_Resource}
}

Nobel-winning Economist Robert Lucas explained that “Once you start thinking about exponential growth, it’s very hard to think about anything else.’’ I agree! 

\subsection{Resisters}

Of course, not every research field has yet transitioned to FR. Possibly, researchers in such fields are not paying attention; or maybe 
adopting a new paradigm \-- the Frictionless Reproducibility triad \--  is just too 
out of their grasp; or maybe they are proud to be preserving ways of the past.

Triad-resisters ought to take stock of what empirical machine learning has achieved in the last decade, and note that while there were some good ideas in machine learning, the rapid development of those ideas, and their outsized impact on intellectual life, was turbocharged by the transition to Frictionless Reproducibility and the emergence of 
the FRX research superpower. If the empirical machine learning community were operating by the older standard of In-Principle Reproducibility -- which many fields still follow, because it’s all they know -- EML might well be today stuck at long-ago performance levels. 

The choice of a research community to resist FR is a choice by that community to be eclipsed as other 
communities who do embrace this practice grow and spread. 
Complacent established researchers may be comfortable with this, or maybe they just never thought about it.

\subsection{Revolutions}

The crossover to FR upends our expectations about research and researcher behaviors, in two ways.

\subsubsection{Computational Epistemology.} 
\label{sec-Epistemology}

In the `comprehensive FR' regime, where every computation can be reproduced, 
adherent researchers no longer depend on hypotheticals. They instead limit themselves to documenting facts:
\begin{description}
\item [What:] a specific workflow does;
\item [On:] a specific publicly available dataset;
\item [According to:] a specific task performance measurement;
\item [Using:] publicly available code. 
\end{description}
This is an extremely limited agenda for research discussion! In exchange for the limitations, one gets two benefits:
\begin{description}
\item [Epistemic modesty.] The research literature is transparent about what is being asserted and under what conditions. 
\item [Full computational reproducibility. ] The research literature becomes a transcript of actual code executions, and so is, simply, true.
\end{description} 

This approach to epistemology is completely a creature of the Frictionless Reproducibility triad and would make no sense without it. 
One is only making claims that are immediately verifiable because of the assumed [FR-1]+[FR-2]+[FR-3] setting. 
This intertwining  is spelled out in the table below:

\begin{table}[htp]
\begin{tabular}{| l | c | c |}
\hline
Facet Documented & Facet explanation & Triad Elements Involved \\
\hline
What: & Specific Workflow Does & [FR-1]+[FR-2]+[FR-3] \\
On:     & Specific, Publicly Available, Dataset & [FR-1: Data Sharing] \\
According to: & Specific Task Performance metric & [FR-3: Challenges ] \\
Using: & Specific Publicly Available Code & [FR-2: Code Sharing ]\\
\hline
\end{tabular}
\caption{Epistemology and FR Components}
\label{tbl:Epistemology}
\end{table}

What is obviously gained is efficiency in article composition, and efficiency in article evaluation. 
Less obvious is that {\it this epistemology enables emergence of an FRX}, which
is a research community exchanging  
digital research artifacts rather than more nebulous entities.

What is potentially gained compared to traditional scientific publication is unprecedented virality. 
In EML, some articles showing a fertile new architectural concept breaking some performance barrier 
can garner massive impact quickly (eg. ResNet, Transformers), by being rapidly incorporated in thousands of works
of other community members.
 
Look at the ranking of top journals sciencewide, ranked by accepted citation metrics. Over the last decade, coincident with the maturation of the data science components [FR-1]+[FR-2]+[FR-3], the journal impact leaderboard, long heavy with incumbent biology, medicine, and general science journals, has been disrupted; 3 of the top 10 journals are today from engineering fields that follow this paper layout\footnote{\url{https://scholar.google.com/citations?view\_op=metrics\_intro\&hl=en}, Accessed September 29, 2023. The new entrants are IEEE Conf. CVPR, NeurIPS, ICLR. Their new peers are Nature, New Eng. Journ. Medicine, Science, Lancet,Adv. Materials, Cell }. Ten years ago, these venues were nowhere to be seen in the journal leaderboard. A true revolution.

It's easy to see why. Traditional science epistemology was heavy with counterfactuals. Authors discussed consequences of some hypothetical intervention on hypothetical measurements and compared those with observed measurements. In fact, over the last century, human intelligence was more or less equated with the articulate use of counterfactuals, conditionals, and hypotheticals in discourse.  The new context removes the burden of articulate deployment and parsing of counterfactuals.

\subsubsection{Research Mindset.} 
There is also a `disruption’ in inter-researcher professional discussion. 
When I observe today's younger data scientists, 
I note that their typical interactions exchange information 
about their information technology stacks. Typical questions include:

\bitem
\item What's your package name?
\item What's your URL? QR Code?
\item Is package $<X>$ on your  stack? ($<X>$varies from conversation to conversation)
\eitem 
More generally, they are often asking for details underlying the FR Triad: 
They want to know are you Data Sharing? Using some new shared dataset I haven’t previously heard of? are you Code Sharing? Using some new shared codebase I haven’t previously heard of? Is there some new specific numerical performance measure? Is there a challenge that just dropped? How can I get single-click access to the work you are describing?
These are all signs that {\bf these researchers are participants in an FRX} \-- a research community exchanging  
digital research artifacts.

Traditional intellectual discourse was heavy with conditionals and counterfactuals – i.e. thought experiments. 
Computation was so difficult that the ability to run mental simulations of what computations might deliver, if we could conceivably do them, was a sign of real intellectual penetration. 
Today it’s a big turn-off to propose thought experiments and impossible hypotheticals like `you could potentially try to do such-and-such’.  One might as well suggest a fly-by of Venus.

The currency of discourse today is Frictionless Replications. 
Researchers are thinking: {\sl Can’t I just try this now?}

\section{Actions Readers can take Immediately}

Here are some action items implied by the argument so far.

\subsection{Exemplify Frictionless Reproducibility.}

In line with the spirit of the times, try to make your own research work capable of viral spread among data scientists. You can best do this by having it be single-click reproducible.

Suppose you are a mathematical scientist who thought of your job as mathematically describing a data science procedure and creating a theorem to probe its behavior. That approach worked for my generation, but it runs afoul of the `URL or it didn’t happen’ mindset of modern life. Today, you might also implement your tool computationally, share code, evaluate it on shared data, and make formalized quantitative comparisons of your work with standard baselines. Implicitly in each research product you can be either participating in an existing challenge, or creating a new reproducible challenge problem and proposing its first entry. If there are today no relevant datasets or baselines, you can productively work to create them.  Creating a dataset or challenge can have much more impact than merely theorem-izing a specific tool. If you don’t immediately see a way to integrate the three elements [FR-1]+[FR-2]+[FR-3] into your work, sense this as a puzzle to unlock an opportunity.


Alternatively, suppose you work in a non-mathematical field, let’s say in medicine, that does not currently use one or more legs of the triad. Work to change this, for example instituting the sharing of data and code and documenting performance. Such work can be visionary and foundational, and there can be grants to support it. This has been proven for example by the Nightingale Open Science project. Medicine is just one example; this pattern can repeat throughout science and technology.

\subsection{Understand the rapidly changing intellectual climate}

The `AI Hegemony' Narrative states that hegemons `own’ brutal scaling, that brutal scaling is the only way, and that therefore hegemons own AI and own the future, forever. 

This depressing vision leaves no room for ideas, for academia, or for ‘the little guy’.  The viewpoint I’m offering you makes clear the self-serving nature of the hegemon narrative. The hegemon CEOs don’t care about the effects of their dispiriting narrative on the psychology and motivation of budding data scientists; but they do care about promoting stock prices. Hence: ‘Owning the future, forever.’

The `AI Singularity' and `AI Doom' narratives can be picked apart similarly. Just keep in mind that the last decade has not been the decade of AI; it has been the data science decade, capping a civilizational project to deliver smartphones to everyone everywhere, supported by a staggering capital investment in compute and communications infrastructure. 

Data science is ubiquitous, but sort of invisible; it has no corporate PR, and no science fiction fanbase. Data Science is not a threat to life on earth, and so lacks the dark charisma of AI. Data Science makes life better; when I think back to the struggles required by computers and data analysis 40, 30, 20, or even 10 years ago, I can see this clearly, in my own life. But no one is being paid to remind me of this, or to scare me about what Data Science may do tomorrow.

\subsection{Rapid changes create opportunities. Chase them down!}

\subsubsection{Research opportunities}

{\it Dataset Bias.} Are the shared public databases we have a reasonable representation of the
dataset universe, or are they in some way skewing our view of what's important?
Are there ways to better incentivize Datafication? Data Sharing? Are there negative consequences
of this approach, perhaps ones we can guard against or mediate?

{\it Availability Bias.} I perceive a heavy bias of researchers towards certain popular approaches over others;
often this is because of Code Sharing. Thus, deep learning is considered `first priority’ by today’s empirical ML researchers whereas, traditional tree-based prediction methods like CART and random forests are considered deprecated, perhaps because the shared code resources for deep learning became more polished and more convenient. For many non-image, non-language so-called `tabular' datasets, the tree-based methods make more intuitive and practical sense. But if few researchers want to use them 
in challenge entries, we face ignorance about whether those tools work, or how they compare.

{\it Challenge Bias.} In the new regime, we have groups of researchers competing to improve performance in task performance on shared public databases.  Are there predictable consequences of this approach, perhaps negative ones we hadn’t thought of? Are there novel computational tools to  ward off the negative effects, or adjust for them?

{\it Challenge Design.} How do we design better Challenges?

{\it Exchange Design.} How do we entice a group of researchers to participate in our challenge?
What practices work and don't work?  How do we ignite a group of researchers into a
dynamic, productive FRX? How do we catalyze FRX takeoff?

{\it Scaling Brutality.} In the new regime, we see a strong commitment to brutal scaling-up of datasets and models. If this were truly `the only way’, what does it say about the underlying prediction task? Traditional mathematical viewpoints about curse of dimensionality and rates of convergence in high-dimensional prediction don’t seem to have been carefully compared against the empirical ML/brutal scaling experience. Can they be reconciled? If not, do we need an entirely new basis for high-dimensional statistics?

{\it Resource Blindness.}  A particular feature of the new regime is that doing things just about as well but dramatically more efficiently is typically not at all valorized. Have brutal scaling and blind focus on performance obscured tremendous efficiencies lying in plain sight? What would the more efficient methods be? How could uncovering them be valorized effectively?

\subsubsection{Teaching Opportunities}

{\it Push Back against Faulty Narratives!}  This paper exposes a clash of narratives, between the popular   `AI Singularity’, and `Brutal Scaling' narratives 
and the contrasting narrative that I prefer, the 
little-known but historically grounded ‘Data Science finally takes hold’.  The AI Singularity narrative promotes powerlessness and is demoralizing to the young.  
Data science instruction can proclaim an accurate and hopeful counter-narrative.

\begin{table}
\fbox{
\parbox{\textwidth}{
\centerline{\bf Narrative: Data Science Takes Hold}

\begin{description}
\item{(a)} a remarkable confluence of developments took place in the last decade, creating a powerful information technology facilities stacked on a global compute, storage, and communications resource.  

\item{(b)} because of these new facilities, the global economy is going through a multiple-decades-long total digital transformation, which will gradually change everything: business, government, education, science.

\item{(c)} in particular, there are many exciting new careers and opportunities adapting these technologies to human needs; in these careers we deploy these new facilities to develop new models and processes.

\item{(d) }The data science principles [FR-1]+[FR-2]+[FR-3] increasingly become the go-to paradigm for research-based model building. Let’s learn this paradigm, learn some of its history and successes, and follow it in our own careers.
\end{description}
}
}
\end{table}

This narrative will better equip students for the future and induce more respect of students for the material they are learning and for the value of their own educations. It can be injected everywhere in academic data science curriculum.

\noindent
{\it New Advanced Courses.}  In Fall 2023, Xiaoyan (XY) Han and I are teaching Statistics 335 at Stanford on the Challenge Problems Paradigm, its components and it consequences, both in ML and elsewhere. As with other 300-level courses in my department, this course may have PhD students and ambitious masters students enrolled. I plan to engage students in many of the issues discussed in this paper, in particular, to thinking about the history to date of this approach, some of its greatest achievements and some variations and modifications. 

\noindent
{\it New Undergraduate Courses.} The game has changed for data-driven science, but actual instruction has not adapted. The National Academies conducted a study \cite{national2018envisioning} of the undergraduate data science curriculum in 2017-2018\footnote{\href{https://www.nationalacademies.org/our-work/envisioning-the-data-science-discipline-the-undergraduate-perspective}{NASEM: ``Envisioning the Data Science Discipline''}}; this already looks in need of severe revision. Many of today’s students simply wouldn’t be able to get value from statistics courses taught the way they were taught before 2010, yet some courses at some institutions haven’t really adapted.  The profusion of publicly shared datasets and data analysis scripts, along with on-line teaching resources such as youtube videos, has already changed the expectations of students, about what they want to learn and how. Probably we don’t yet agree on what comes next, but our students are in a completely new mindset and completely new approaches are needed. 

\subsubsection{Startup opportunities}

Frictionless reproducibility is here, now; in the future we will see more and more research fields following its discipline, meaning there is an opportunity for services and tools to support and build upon the triad [FR-1]+[FR-2]+[FR-3], including outside the context of empirical machine learning. To me the big opportunity lies in making some of the predictions listed below, come true.

\pagebreak

\section{Predictions}

This paper began life for a session on ‘The Next 50 years of Data Science’ at the 2023 Joint Statistical Meetings.
Implicitly, we were being asked for predictions, 
Here are several. See also \cite{donoho201750}.

\subsection{Computation on Research Artifacts (CORA)}

Frictionless Reproducibility in computational research is now essentially here. As a side effect this makes available a vast array of digital research artifacts generated by earlier research.   Soon, those byproducts will be viewed as digital gold.

Here’s how. In many research disciplines, reports spotlight tables which constitute the main deliverables of the work being presented. Such tables bring together chosen data and algorithms and specific task performance metrics.  In methodological work, they can often be viewed as what an EML researcher would call a `mini-challenge leaderboard’; I am speaking here inclusively, broadly, outside the EML context, and allow an enlarged definition of task and task performance. 

In the near future, researchers will understand that future researchers, digesting the given research work, would wish to directly probe the computations that generated the numbers in the researcher’s tables, as a way of engaging with the researcher’s result and possibly building upon it. Whole communities will expect their researchers to expose the data and workflow that produced the surfaced numerical results. 

New tools will soon enough become available to automatically enable frictionless access to the underlying digital artifacts that went into any relevant table in a study. This will happen seamlessly, as part of ordinary research and publication; standard research computing and publication computing environments, with little or no effort, will automatically expose the underlying data, code and performance measurements that delivered these results. 

Exposing such artifacts enables other data scientists who read a published report to build upon the research that has been done in that report: to first reproduce and, later, to possibly modify some underlying algorithms or some underlying observables and recompute the table, thereby producing a new research project.  

This enters us all into a new era, where {\bf computing on} the digital {\bf research artifacts} created by previous research computing (for which we adopt the label CORA) allows us to write algorithms to inspect and generalize the workflows used in previous research, and thereby algorithmically obtain new research results.  We will begin to naturally think and operationalize at a new level as follows: “Take project X’s workflow and everywhere replace algorithm A with algorithm B, call the result  `project Y’s workflow'. Execute the full workflow of project Y, computing all the same observables as you did for Project X.” 

As a vision, CORA has been clear to many data scientists for some time. The Quantitative Programming Environment Mathematica and its associated notebook system delivered some of this functionally already, at least 30 years ago. Stephen Wolfram in several venues has evoked his vision of computing on anything, broad enough to include CORA\footnote{To share the `vibes', I change one word in this `typical' Wolframism: `... we are finally now getting into a position where we can take all that {\sl [RESEARCH]} that we as humans have accumulated in our civilization, and encapsulate it in an active computational form. And when we do this, we make it possible to dramatically extend and generalize all our achievements so far.'  (almost-quoted-from) Stephen Wolfram, {\it On the Quest for Computable Knowledge, 2017.} }.

CodaLab\footnote{\url{https://worksheets.codalab.org/}},\footnote{\url{https://codalab.lisn.upsaclay.fr/}} is a platform for running challenge problem contests that can inherently perform CORA, for certain specific tasks (e.g. swap the public test dataset that was used by a certain contest entry, for a certain private dataset sequestered from public view, and re-evaluate specific performance metrics after the swap).\footnote{The NMRBox system discussed in Section \ref{sec:FREverywhere} has some analogous features}.

My prediction is that CORA will inevitably emerge and grow in scope, to outstanding effect, increasing the pace of scientific advance.  My thinking derives from  (practically unknown) papers by the author and Matan Gavish \cite{gavish2011universal,gavish2012three} in which we described the notion of {\it Verifiable Computational Results} (VCR), 
and three dream applications. 

VCR, where implemented, ensures that digital artifacts emerging during a computation are automatically captured and made frictionlessly available to future researchers; and dream applications exploit these artifacts to automatically generalize upon earlier work (eg swap this computational method for the one originally used in prior research, and show me the new table that results) or automatically evaluate earlier research (eg bootstrap this entry in that previously published table and show me the histogram that results).

It seems to me that full implementation of CORA, when it happens, will be a transformative achievement! Also, that we as a scientific computing world are headed towards that achievement, and its fulfillment dates {\it the actual moment of the reproducibility singularity}. 

\subsection{CORA and the `End Times'.} 

Kurzweil calls attention to a very dramatic moment, coming soon – yielding priority to IJ Good,  {\sl Speculations concerning the first ultraintelligent machine}, 1965 \cite{good1966speculations}:
\begin{quotation}
Let an ultraintelligent machine be defined as a machine that can far surpass all the intellectual activities of any man however clever. Since the design of machines is one of these intellectual activities, an ultraintelligent machine could design even better machines; there would then unquestionably be an “intelligence explosion,” and the intelligence of man would be left far behind. Thus the first ultraintelligent machine is the last invention that man need ever make.
\end{quotation}

Good was an amazing intellectual\cite{good1983good}, who served as Alan Turing’s research assistant and who `Kept Bayes Alive’ during its exile from academic favor\cite{Neapolitan2008}.  As you can see, he was a stylish writer and his formulation is charismatic and compelling. It’s easy to see why Kurzweil latches onto it and how it might inspire “AI Doom” influencer threads; Eliezer Yudkowsky coined\footnote{\url{https://www.lesswrong.com/posts/tjH8XPxAnr6JRbh7k/hard-takeoff}} the notion of a “hard takeoff” or FOOM\footnote{\url{https://www.lesswrong.com/tag/ai-takeoff}}, where machine intelligence starts to expand its own powers hyper-exponentially. Good seems to anticipate hard takeoff.

Back to Earth. Let’s acknowledge that, in certain ways, performance on challenges is inevitably and automatically going to improve markedly post-transition to FR. 
Indeed, it is a simple matter to take any given frictionlessly reproducible workflow and modify it by trial and error to produce a new workflow that may improve it. Consider two meta-operators on workflows:

\begin{description}
\item [Transformation 1: ] {Wrappers\;} 
\bitem
\item Adopt a given existing workflow as a starting point. 
\item Prepend a new workflow element to preprocess the inputs
\item Append a new workflow element to postprocess the outputs
\item Obtain a new workflow
\eitem
\item [Transformation 2: ] { Tweaks\;} 
\bitem
\item Adopt a given existing workflow as a starting point. 
\item Assuming this workflow exposes internal hyperparameters that were previously unexplored:
\bitem 
\item Systematically vary those unexplored hyperparameters
\item Extract the optimal hyperparameter combination
\item Obtain a new workflow
\eitem
\eitem
\end{description}
Much of the EML research activity we have seen in recent years 
reports discoveries of 
successful applications of these two strategies,
in many cases adding new layers/gadgets to a multilayer architecture (Transformation 1); 
or modifying an existing architecture by, let’s say, adding extra width (Transformation 2).  
The new models produced by such meta-operations can then be evaluated 
robotically to find one which improves performance. 

This has been automated in empirical machine learning (eg. with famous systems like 
AutoML \cite{hutter2015automatic,hutter2019automated,AutoMLSurvey2022}, and Vizier \cite{golovin2017google,song2023open}; 
the process points in the direction of Jack Good’s vision. We avoid human guesswork and blind luck to get good models; hyperparameter search is done to a certain degree robotically.

These meta-operators involve operations on workflows \cite{drori2021alphad3m} -- digital artifacts produced by previous research computing. Hence these are instances of what has earlier been discussed and labelled CORA. In any field that invests in the FR regime, the digital artifacts are there, as a side effect of publication, available frictionlessly to be exploited. It is just a matter of adopting the mindset to exploit the artifacts to do future research. 

As individual fields invest in FR, they will transition to CORA; the transition may even enable algorithmic improvement of models
and produce a hard takeoff of model performance. 
Maybe the last decade's steep increase in performance in EML is an
approximation to hard takeoff; the approximation manifesting the 
 `still-slightly-frictioned',
`still-human-handwork-was-involved'  `CORA-was-still-in-its-infancy' 
situation over the last decade.

Will CORA make machines hyper-intelligent by itself? 
No. All we will get via automated application of strategies like those
listed here is incremental bumps to performance. 
We will painlessly and rapidly mine out the value of each human-supplied idea. 

Peter Huber, in his 2012 book {\it Data Analysis: What can be learned from the last 50 years} \cite{huber2012data} contrasts strategy (high-level wisdom) and tactics (low-level, step-by-step). Using his terminology, we predict that, as FR takes hold in a field, and CORA becomes ubiquitous, humans will increasingly focus on strategic choices about model architectures and types of hyperparameters; while increasingly, information technology will automate the tactical details of finding the best instances within those strategic constraints.

\subsection{What's Next}

Certain features of research life going forward from that point are very clear. 

\bitem
\item {\sl Irreversibility.}  The transition to FR is a one-way transition: the wishes underlying  [FR-1]+[FR-2]+[FR-3] have been around as long as there has been data-driven science; the ability to fulfill those wishes is now provided by a complete technology stack which came into existence to service the global communications/computation infrastructure. That infrastructure is here to stay.
The desire for FR won’t go away; the ability to offer FR won’t go away either. 

\item {\sl Acceleration.} Research fields that have transitioned to the Frictionless Reproducibility regime will progress even faster in the future as the habits get even more ingrained, producing even higher quality  with ever higher velocity. Further, a community of researchers working together according to this pattern will recognize and seize new opportunities as a result.

\item {\sl Stagnation.} Some fields won’t or can’t practice mature forms of the three data science initiatives we have discussed. Barriers to transition can include: inhibitions against data sharing, for example, because of confidentiality restrictions; inhibitions against code sharing, for example because of proprietary restrictions; inhibitions against prediction challenges, for example because of 
 a preference for theoretical derivations disconnected from empirical measurements.

Such fields won’t transition to a Frictionless Reproducibility regime. They will be noticeably lagging behind in rate of progress. Those fields may soon enough be recognized as backwaters.

\item {\sl Resource Allocation.}  Funding agencies will recognize that they get higher rates of return on investments in research communities which have transitioned to Frictionless Reproducibility; they will invest more heavily in such fields over time. Research talent will follow the dollars. The separation between fields, initially from reproducibility practices,
 will then involve funding, compounding the contrast between accelerating and stagnating fields.

\eitem


Some other features of our future are less obvious. 

\bitem
\item{\sl Opacity.}  A key point made by Kurzweil in “The Singularity is Near” is that progress “on the other side” of “The Singularity” will be so amazing, we can’t even imagine what will be accomplished.  This seems fair to me. Seniors among us became used to the idea that certain research topics progress slowly or not at all; 
are best learned by dint of long study and philosophical reflection; and research contributions happen on the time scale of entire careers.  
Minds trained in such a regime might not adjust to how rapidly things will soon be happening.

Of course some research topics are, so far, untransformed by the three data science initiatives. But eventually some will be so transformed, and research in such areas will rapidly move far beyond what experts of earlier generations would have thought possible, or even conceivable.  Others may be, essentially forgotten. Hence, the Reproducibility Singularity is indeed opaque.

\item{\sl Amnesia.} The opacity of the Reproducibility Singularity works in reverse.  Much of the work pre-reproducibility transition will be forgotten. Listening to the young makes this clear. 

Current Stanford Statistics Ph.D. student Apratim Dey told me about a Stanford CS Class on ``Distribution Shift" in Spring 2023. In that class, the initial course reading was a more or less traditional 
academic Statistics journal article from 20 or so years ago that established the field (more or less). That paper posed a generative probabilistic model for the distribution shift problem and performed a formal analysis using it. Later in the term, papers came from leading CS conferences and followed a more empirical bent, using shared data and code.

The CS students in this course apparently were not fans of the academic statistics paper, although they were told it was fundamental. 
In discussing this paper, they pursued critiques along these lines:
\bitem
\item  It's based on a theoretical generative model
\item I don't know how to tweak or improve a theoretical generative model
\item I don't know if this generative model describes any real data
\item I don't know if the phenomena it describes occur on any real data
\item If I do research on real data, I can share results with others, publish in a CS conference.
\item If I instead do research with a generative model, as is done here: I have no immediate outlet for my work, no readership, no follow-up. It could take years and years to publish in a traditional stats journal. 
\eitem

Maybe these are just comments about one paper; but possibly they are comments about papers of this type. 
Namely, papers that don’t hew to today’s empirical data science style, which involves building data and code and empirical performance
on the work of others. 
The CS students don’t see the older foundational research as connecting to today’s frictionlessly reproducible regime; 
it doesn't offer the opportunity for single-click adoption and code re-execution by students. 
Engaging with such non-conforming work disrupts the students' state of `flow’; 
hence such papers are seen as frustrating. Such work confronts students, in their view, with gotchas and obstacles. Also they don’t see how
engaging with it allows them to efficiently progress to publication 
in the outlets they prefer. Such pre-FR work  offers friction, not fulfillment.

This younger CS generation is steeped in the recent empirical machine learning literature, much of which is powered by FR/FRX; it 
seemingly doesn’t warm to the style of mathematically powered discourse practiced previously, which is often based on hypotheticals and counterfactuals;
just maybe, it doesn’t see the need to learn how to engage with such work. 

We may be starting a rapid de-skilling transition, in which new researchers initially don’t want to, and later simply can’t onboard 
the lessons of symbolic analysis of generative models.  

\eitem

\section{Conclusion}

A data-science-driven phase transition, crossing a kind of singularity, is happening now. 

Not the `AI singularity' ---  the moment when AGI is achieved  --- this hasn’t happened. 
Rather, some fields in computation-driven research are approaching a Reproducibility Singularity, 
the moment when all the ingredients of frictionless reproducibility (FR) of computational research 
are assembled in a package that makes it, essentially, immediate to build on the work of 
other researchers on an open, global stage. 
A research community working fully within this regime 
creates and exchanges digital artifacts following new models of scientific
publication and epistemology;  this is a research superpower
enabling strikingly rapid progress.

This new regime has been best approximated within the field of empirical machine learning (EML), 
which indeed has made very rapid progress over the last decade in the fields of computer vision and natural language processing. 

The FR regime is best thought of, not as  `property' of EML, but 
instead as a more general achievement of dedicated Data Scientists working for decades, 
working in many research disciplines, including those where empirical machine learning and AI 
historically plays little role.  

Today’s `AI Singularity' narrative -- that the last decade was the decade where AI was achieved -- is premature; 
this was instead the decade when data science matured, where a global computing and communication infrastructure came 
together, and frictionless delivery of research artifacts emerged, enabling computational science as a 
solid new avenue to scientific validity.

The next generation of data scientists can leverage this understanding in building 
their own careers to identify interesting questions to work on and to make their research impact more viral.


\section*{Acknowledgements}

I would like to thank Genevera Allen of Rice University for organizing the session ‘The Next 50 Years of Data Science’ at the 2023 Joint Statistical Meetings in Toronto and giving me the opportunity to participate.  I would also like to thank my fellow committee members on the NASEM Committee on Reproducibility and Replication (Harvey Fienberg, Chair), from whom I learned a great deal about the global, sciencewide nature of the changes going on in science and the advances in reproducibility. Also, I thank my fellow members on the NASEM study on Envisioning the Data Science Discipline: The Undergraduate Perspective(David Culler, Chair) for making me aware of the many currents of change in Data Science education.

Brian Wandell (Stanford) provoked the title. In alphabetical order,  Milad Bakhshizadeh(Stanford), Apratim Dey (Stanford), Xiaoyan Han (Cornell), Vardan Papyan (Toronto),  Elad Romanov (Stanford) and Yu Wang (Stanford) made many helpful comments. This is closely related to work in progress by the author and Matan Gavish (Hebrew University); I also want to acknowledge Victoria Stodden (University of Southern California) for many conversations on reproducibility over the years.

\bibliographystyle{ieeetr}

\bibliography{Singularity}

\newpage

\begin{table}
\fbox{
\parbox{\textwidth}{
\centerline{\bf Explain it like I'm 5 years old}

\begin{description}

\item{(5yo)} Things are happening awfully fast these days! I frequently see advances in data  processing --
text, video, tabular data -- coming at a rate I hadn't been expecting to see.
\item{(5yo)} This makes it look like there's a superpower out there. Maybe it's AI?
\item{(Me)} No, actually it's data science. 
\item{(5yo)} What's the data science superpower?
\item{(Me)} Okay, on the fingers of one hand: 
\begin{description}
\item{(1)} In the last decade, frictionless services became available thanks to the modern smartphone information ecosystem
\item{(2)} Those frictionless services were applied by Scientists and Technologists to Data Sharing, Code Sharing and Challenges
\item{(3)} Some communities of researchers started frictionlessly sharing research artifacts -- code, data, results -- and building on each others' work.
\item{(4)} Involved research communities are progressing much faster. It's night and day.
\item{(5)} Soon it will {\it really} speed up, as scientists write computer programs to exploit all the digital artifacts
from previous research.
\end{description}
\item{(5yo)} But wait, I'm still reading everywhere about AI.
\item {(Me)} AI  is one of those communities where people are working this way, that's why you're hearing so much news about it.
\item{(5yo)} But I heard that there's a singularity, that AI is going to unemploy everybody and may kill us all.
\item {(Me)} There is a singularity (= new superpower) but not in AI. It's a superpower allowing us to do research faster and better.
It's not dangerous. You'll just be hearing more and more frequently about advances, because research goes faster and faster.

\end{description}
}
}
\end{table}

\begin{table}[htp]
\centering
\begin{tabular}{| l | l |}
\hline 
Abbrevation & Meaning\\
\hline
AI        & Artificial Intelligence \\
BYODC & Bring-your-own-data Challenge\\
CASP & Critical Assessment of Protein Structures \\
CORA & Computation on Research Artifacts \\
CS & Computer Science \\
CVPR & Computer Vision and Pattern Recognition \\
DARPA & Defense Advanced Projects Research Agency \\
EML & Empirical Machine Learning \\
FOOM & Hard AI Takeoff \\
FR & Frictionless Reproducibility \\
FRX & Frictionless Research Exchange \\
GPU & Graphics Processing unit \\
IPR & In-principle reproducibility \\
ICLR & International Conference on Learning Representations\\
NASEM & National Academies of Science, Engineering, and Medicine \\
NeurIPS & Neural Information Processing Systems \\
NIST & National Institute of Standards and Technology \\
OSS & Open-Source Software \\
PMC & Proprietary-Methods Challenge \\
RCS & Reproducible Computational Science \\
\hline
\end{tabular}
\caption{Abbreviations used in this article}
\end{table}

\begin{table}[htp]
\centering
\begin{tabular}{| l | l |}
\hline 
Label & Meaning\\
\hline
AI Doom     & Artificial Intelligence will wreak terrible destruction\\
AI Extinction &  Artificial Intelligence will extinguish humanity\\
AI Hegemony & Artificial Intelligence is effectively owned by some few Internet Hegemons\\
AI  Singularity       & Artificial Intelligence is soon to disrupt human affairs  \\
Brutal Scaling & Massive data and compute are sufficient and necessary for performance \\
Emergent Superpowers & Unprecedented powers that emerge under `just right' conditions \\
Hard Takeoff & Artificial intelligence suddenly improves exponentially and dominates us \\
\hline
\end{tabular}
\caption{Narrative Labels}
\end{table}

\end{document}